



\font\bigbold=cmbx12  \font\tenfrak=eufm10  
\font\sevenfrak=eufm7 \font\fivefrak=eufm5  \font\tenbb=msbm10
   \font\sevenbb=msbm7   \font\fivebb=msbm5
\font\tensmc=cmcsc10


\newfam\bbfam
\textfont\bbfam=\tenbb
\scriptfont\bbfam=\sevenbb
\scriptscriptfont\bbfam=\fivebb
\def\Bbb{\fam\bbfam}

\newfam\frakfam
\textfont\frakfam=\tenfrak
\scriptfont\frakfam=\sevenfrak
\scriptscriptfont\frakfam=\fivefrak
\def\frak{\fam\frakfam}


\def\pagewidth#1{\hsize= #1}
\def\pageheight#1{\vsize= #1}
\def\hcorrection#1{\advance\hoffset by #1}
\def\vcorrection#1{\advance\voffset by #1}
\newif\iftitlepage   \titlepagetrue               
\newtoks\titlepagefoot     \titlepagefoot={\hfil} 
\newtoks\otherpagesfoot    \otherpagesfoot={\hfil\tenrm\folio\hfil}
\footline={\iftitlepage\the\titlepagefoot\global\titlepagefalse
           \else\the\otherpagesfoot\fi}
\def\abstract#1{{\baselineskip=12pt\parindent=30pt\narrower\noindent #1\par
\baselineskip=16pt plus 1pt minus 1pt}}

\newcount\notenumber  \notenumber=1
\def\note#1{\unskip\footnote{\baselineskip=10pt$^{\the\notenumber}$}
{#1}\global\advance\notenumber by 1
\baselineskip=16pt plus 1pt minus 1pt}
\def\smc{\tensmc}


\global\newcount\secno \global\secno=0
\global\newcount\meqno \global\meqno=1
\global\newcount\appno \global\appno=0
\newwrite\eqmac
\def\romappno{\ifcase\appno\or A\or B\or C\or D\or E\or F\or G\or H
\or I\or J\or K\or L\or M\or N\or O\or P\or Q\or R\or S\or T\or U\or
V\or W\or X\or Y\or Z\fi}
\def\eqn#1{
        \ifnum\secno>0
            \eqno(\the\secno.\the\meqno)\xdef#1{\the\secno.\the\meqno}
          \else\ifnum\appno>0
            \eqno({\rm\romappno}.\the\meqno)\xdef#1{{\rm\romappno}.\the\meqno}
          \else
            \eqno(\the\meqno)\xdef#1{\the\meqno}
          \fi
        \fi
\global\advance\meqno by1 }


\global\newcount\refno
\global\refno=1 \newwrite\reffile
\newwrite\refmac
\newlinechar=`\^^J
\def\ref#1#2{\the\refno\nref#1{#2}}
\def\nref#1#2{\xdef#1{\the\refno}
\ifnum\refno=1\immediate\openout\reffile=refs.tmp\fi
\immediate\write\reffile{
     \noexpand\item{[\noexpand#1]\ }#2\noexpand\nobreak.}
     \immediate\write\refmac{\def\noexpand#1{\the\refno}}
   \global\advance\refno by1}
\def\semi{;\hfil\noexpand\break ^^J}
\def\nl{\hfil\noexpand\break ^^J}
\def\refn#1#2{\nref#1{#2}}

\def\cmp#1#2#3{{\it Commun. Math. Phys.} {\bf {#1}} (19{#2}) #3}
\def\jmp#1#2#3{{\it J. Math. Phys.} {\bf {#1}} (19{#2}) #3}
\def\ijmp#1#2#3{{\it Int. J. Mod. Phys.} {\bf A{#1}} (19{#2}) #3}
\def\mplA#1#2#3{{\it Mod. Phys. Lett.} {\bf A{#1}} (19{#2}) #3}
\def\pl#1#2#3{{\it Phys. Lett.} {\bf {#1}B} (19{#2}) #3}
\def\np#1#2#3{{\it Nucl. Phys.} {\bf B{#1}} (19{#2}) #3}
\def\pr#1#2#3{{\it Phys. Rev.} {\bf {#1}} (19{#2}) #3}

\def\prD#1#2#3{{\it Phys. Rev.} {\bf D{#1}} (19{#2}) #3}
\def\prl#1#2#3{{\it Phys. Rev. Lett.} {\bf #1} (19{#2}) #3}


\expandafter\ifx\csname amssym.def\endcsname\relax \else\endinput\fi
\expandafter\edef\csname amssym.def\endcsname{%
       \catcode`\noexpand\@=\the\catcode`\@\space}
\catcode`\@=11
\def\newsymbol#1#2#3#4#5{\let\next@\relax
 \ifnum#2=\@ne\let\next@\msafam@\else
 \ifnum#2=\tw@\let\next@\bbfam@\fi\fi
 \mathchardef#1="#3\next@#4#5}
\def\mathhexbox@#1#2#3{\relax
 \ifmmode\mathpalette{}{\m@th\mathchar"#1#2#3}%
 \else\leavevmode\hbox{$\m@th\mathchar"#1#2#3$}\fi}
\def\hexnumber@#1{\ifcase#1 0\or 1\or 2\or 3\or 4\or 5\or 6\or 7\or 8\or
 9\or A\or B\or C\or D\or E\or F\fi}
\edef\bbfam@{\hexnumber@\bbfam}
\csname amssym.def\endcsname

\pageheight{23cm}
\pagewidth{15.5cm}
\hcorrection{-2.5mm}
\magnification \magstep1
\baselineskip=16pt plus 1pt minus 1pt
\parskip=5pt plus 1pt minus 1pt


\def\smc{\tensmc}
\def\frac#1#2{{#1\over#2}}
\def\dfrac#1#2{{\displaystyle{#1\over#2}}}
\def\tfrac#1#2{{\textstyle{#1\over#2}}}
\def\({\left(}
\def\){\right)}
\def\<{\langle}
\def\>{\rangle}
\def\a{\alpha}
\def\b{\beta}
\def\d{\delta}
\def\e{\varepsilon}
\def\s{\sigma}
\def\w{\wedge}
\def\g{{\frak g}}
\def\r{{\frak r}}
\def\h{{\frak h}}
\def\Z{{\Bbb Z}}
\def\Real{{\Bbb R}}

\def\C{{\Bbb C}}

\def\D{{\cal D}}
\def\H{{\cal H}}
\def\ket#1{|#1\rangle}
\def\bra#1{\langle#1|}
\def\phy{{\hbox{\sevenrm phy}}}
\def\pro{\mathop{\rm pr}\nolimits}
\def\tr{\mathop{\rm tr}\nolimits}
\def\Tr{\mathop{\rm Tr}\nolimits}
\def\dim{\mathop{\rm dim}\nolimits}
\newsymbol\ltimes 226E
\def\L{L}     
\def\R{R}     

\def\cs#1{\chi_{_{\rm{#1}}}}
\def\pa{\partial}
\def\SK{{\rm S}_K}

\null
\rightline{DIAS-STP-93-14}
\rightline{August 1993}
\vfill
{\baselineskip=18pt
\centerline{\bigbold On the Emergence of Gauge Structures and
Generalized Spin }
\centerline{\bigbold when Quantizing on a Coset Space}
}
\vskip 30pt
\centerline{\smc David McMullan{\baselineskip=10pt\footnote{$^*$}
{\rm Address after 1.9.93:
School of Mathematics and Statistics, University of Plymouth, Drake
Circus,
Plymouth, Devon PL4 8AA, U.K.}}
and Izumi Tsutsui}
\vskip 20pt
\centerline{Dublin Institute for Advanced Studies}
\centerline{10 Burlington Road}
\centerline{Dublin 4}
\centerline{Ireland}
\centerline{ e-mail: mcmullan@stp.dias.ie\quad tsutsui@stp.dias.ie}
\vskip 40pt
\abstract{%
{\bf Abstract.}\quad It has been known for some time that there are
many inequivalent quantizations possible when the configuration
space of a system is a coset space G/H. Viewing this classical
system as a constrained system on the group G, we show that these
inequivalent quantizations can be recovered from a generalization of
Dirac's approach to the quantization of such a constrained system
within which the classical first class
constraints (generating the H-action on G) are allowed to become
anomalous (second class) when quantizing. The resulting
quantum theories  are characterized by the emergence of a Yang-Mills
connection, with quantized couplings, and new \lq spin\rq\ degrees of
{}freedom. Various applications of this procedure are presented in
detail: including a new account of how spin can be described within a
path-integral formalism, and how on $S^4$ chiral spin degrees of
{}freedom emerge, coupled to a BPST instanton.
}
\vfill\eject


{
\refn\Dirac
{P.A.M. Dirac, \lq\lq Lectures on Quantum
Mechanics\rq\rq\ (Yeshiva, New York, 1964)}

\refn\Jackiw
{R. Jackiw, in \lq\lq Relativity, Groups and Topology
II\rq\rq\   eds. B.S. DeWitt and R. Stora\nl(North-Holland,
Amsterdam, 1984)}

\refn\Kuchar
{K.V. Kucha\v r, \prD{35}{87}{596}}

\refn\MP
{D. McMullan and J. Paterson, \jmp{30}{89}{477, 487}}

\refn\Singer
{I. Singer, \cmp{60}{78}{7}}

\refn\Mackey
{G.W. Mackey, \lq\lq Induced Representations of Groups
and Quantum Mechanics\rq\rq\nl (Benjamin, New York, 1969)}

\refn\Woodhouse
{See, for example,
N. Woodhouse, \lq\lq Geometric Quantization\rq\rq\
(OUP, Oxford, 1980)}

\refn\Isham
{C.J. Isham, in \lq\lq Relativity, Groups and Topology
II\rq\rq\   eds. B.S. DeWitt and R. Stora\nl(North-Holland,
Amsterdam, 1984)}

\refn\LandsmanA
{N.P. Landsman, {\it Rev. Math. Phys.} {\bf 2} (1990)
45, 73}

\refn\LL
{N.P. Landsman and N. Linden, \np{365}{91}{121}}

\refn\MT
{D. McMullan and I. Tsutsui, \lq\lq Functional gauge structures in
gauge theories\rq\rq,\nl in preparation}

\refn\Fadshat
{L. Faddeev and S. Shatashviti, \pl{167}{86}{225} }

\refn\NR
{H.B. Nielsen and D. Rohrlich, \np{299}{88}{471}}

\refn\Varadarajan
{V.S. Varadarajan, \lq\lq Geometry of Quantum
Theory\rq\rq \ vol II\ \nl (Van Nostrand, New York, 1970)}

\refn\Koch
{M. Koch, \lq\lq Quantum mechanics on homogeneous configuration
spaces and classical particles in Yang-Mills fields\rq\rq,
Universit\"at Freiburg preprint THEP 91/19}

\refn\Landsman
{N.P. Landsman, {\it Rev. Math. Phys.} {\bf 4} (1992)
503}

\refn\Ohnuki
{Y. Ohnuki and S. Kitakado, \mplA{7}{92}{2477}; \jmp{34}{93}{2827}}

\refn\Tanimura
{S. Tanimura, \lq\lq Quantum mechanics on manifolds\rq\rq,
\nl Nagoya Univ. preprint DPNU-93-21}

\refn\Perelomov
{A.M. Perelomov, \lq\lq Integrable Systems of
Classical mechanics and Lie Algebras\rq\rq\nl
(Birkh\"auser Verlag, Basel, 1990)}

\refn\Schulman
{L.S. Schulman, \lq\lq Techniques and Applications of
Path Integration\rq\rq\nl (Wiely, New York, 1981);
L. Schulman, \pr{176}{68}{1558};\nl
J.F. Hamilton and L.S. Schulman, \jmp{12}{71}{161}}

\refn\Barut
{A.O. Barut and R. Raczka, \lq\lq Theory of Group
Representations and Applications\rq\rq\nl
(Polish Scientific Publishers, Warszawa, 1977)}

\refn\FS
{L.D. Faddeev, {\it Theor. Math. Phys.} {\bf 1} (1970) 1; \nl
 P. Senjanovic, {\it Ann. Phys.} {\bf 100} (1976) 227}

\refn\David
{D. McMullan, \lq\lq Constrained quantization, gauge fixing and the
Gribov ambiguity\rq\rq, DIAS-STP-92-23 to appear in {\it Commun. Math. Phys}}

\refn\Fadetal
{A. Alekseev, L. Faddeev and S. Shatashvili, {\it J.
Geom. Phys.} {\bf 5} (1989) 391}

\refn\AM
{R. Abraham and J.E. Marsden, \lq\lq Foundations of Machanics\rq\rq\
Second Edition,\nl (Benjamin/Cummings, Reading, 1978)}

\refn\Humphreys
{S.E. Humphreys, \lq\lq Introduction to Lie Algebras
and Representation Theory\rq\rq\nl
(Springer-Verlag, Berlin, 1972)}

\refn\Wong
{S.K. Wong, {\it Nuovo Cimento} {\bf 65A} (1970) 689}

\refn\WZ
{Y.-S. Wu and A. Zee, \np{258}{85}{157}}

\refn\Balachandran
{A.P. Balachandran, G. Marmo, B.-S. Skagerstam and A.
Stern, \nl
\lq\lq Gauge Symmetries and Fibre Bundles: Application to
Particle Dynamics\rq\rq, \nl
Lecture Notes in Physics {\bf 188}
(Springer-Verlag, Berlin, 1983)}

\refn\BB
{F.A. Bais and P. Batenburg, \np{269}{86}{363}}

\refn\Laquer
{H.T. Laquer, in \lq\lq Group actions on manifolds\rq\rq,
{\it Contemp. Math. \bf 36}\nl (AMS, Providence, 1983)}

\refn\TF
{I. Tsutsui and L. Feh\'er, \pl{294}{92}{209}}

\refn\LLA
{N.P. Landsman and N. Linden, \np{371}{92}{415}}

\refn\Nelson
{P. Nelson and L. Alveraz-Gaum\'e, \cmp{99}{85}{103}}

\refn\AB
{Y. Aharonov and D. Bohm, \pr{115}{59}{485}}

\refn\Berry
{M.V. Berry, {\it Proc. R. Lond. } {\bf A392} (1984) 45}

\refn\Wilczek
{F. Wilczek and A. Zee, \prl{52}{84}{2111}}

\refn\Kuratsuji
{H. Kuratsuji, \prl{61}{88}{1687}; \ijmp{7}{92}{4595}; \nl
 H. Kuratsuji and K. Takada, \mplA{5}{90}{1765}}

\refn\Gilmore
{R. Gilmore, \lq\lq Lie Groups, Lie Algebras and Some of Their
Applications\rq\rq\nl
(Willey, New York, 1974)}

}


\dimen1=1.5in       
\dimen2=1in         
\def\classical{\mathop{{\tt CLASSICAL}}}
\def\quantum{\mathop{{\tt QUANTUM}}}
\def\arrowto{\vbox{
  \hbox to \dimen1{\rightarrowfill}}}
\def\arrowsto{\vbox{
   \hbox to \dimen1{\rightarrowfill}
   \kern-10pt
   \hbox to \dimen1{\rightarrowfill}
   \kern-10pt
   \hbox to \dimen1{\rightarrowfill}}}
\def\arrowdown{\hbox{$
     \left\downarrow\vcenter to \dimen2{}\right.$}}

\def\arrowsdown{\hbox{$
     \left\downarrow\vcenter to \dimen2{}\right.\!\!\!
     \left\downarrow\vcenter to \dimen2{}\right.\!\!\!
     \left\downarrow\vcenter to \dimen2{}\right.$}}
\def\harrow#1#2{\smash{\mathop{\arrowto}
       \limits^{\hbox{\rm#1}}_{\hbox{\rm#2}}}}
\def\harrows#1#2{\smash{\mathop{\arrowsto}
       \limits^{\hbox{\rm#1}}_{\hbox{\rm#2}}}}
\def\varrow#1#2{\llap{#1}\arrowdown\rlap{#2}}

\def\varrows#1#2{\llap{#1}\arrowsdown\rlap{#2}}
\def\diagram#1{{\normallineskip=7.5pt\normalbaselineskip=0pt\matrix{#1}}}
\def\stmta{\classical\limits^{^{\displaystyle{\tt EXTENDED}}}
_{_{_{\displaystyle{\tt SYSTEM}}}}}
\def\stmtb{\quantum\limits^{^{\displaystyle{\tt EXTENDED}}}
_{_{{\displaystyle{\tt SYSTEM}}}}}
\def\stmtc{\classical\limits^{^{\displaystyle{\tt REDUCED}}}
_{_{_{\displaystyle{\tt SYSTEM}}}}}
\def\stmtd{\quantum\limits^{^{\displaystyle{\tt REDUCED}}}
_{_{{\displaystyle{\tt SYSTEMS}}}}}
\def\stmte{\quantum\limits^{^{\displaystyle{\tt REDUCED}}}
_{_{{\displaystyle{\tt SYSTEM}}}}}

\noindent \secno=1 \meqno=1

\noindent
{\bf 1. Introduction}
\medskip
\noindent
Yang-Mills
theory provides the theoretical basis for all descriptions of high
energy physics, yet we still know very little about the actual
quantization of such a system. One of the biggest problems is that
the very gauge invariance that makes the theory so attractive
implies that not all components of the gauge field are physical.
This is a familiar problem in electrodynamics where
it is only the transverse components of the photon that have an
independent physical significance.  Based on this observation
one can quantize electrodynamics after reducing it to
the true degrees of freedom\note
{This is, of course, possible only if one is
willing to give up the manifest Lorentz invariance of the
quantum theory.  If not, then even in electrodynamics one has to
deal with an extended system in one way or another.}.
The technical complication in Yang-Mills
theory then being that there is no similar (or at least useful)
classical reduction to the true degrees of freedom, and hence
we are
forced to quantize more than the physical content of the theory.
Clearly, some method is then needed for extracting the physical
sector from
the extended quantum theory.

A systematic approach to this type of problem was initiated by Dirac
[\Dirac] and has become known as Dirac's approach to the
quantization of a constrained system.
Working in a slightly more general context
than the specific example of Yang-Mills theory (for the traditional
quantization of Yang-Mills theory we refer to Ref.\thinspace\Jackiw),
consider a classical system given by an extended phase space
with first class constraints,
$$
\phi_i=0\,.
\eqn\con
$$
Being
{}first class simply means that under the Poisson bracket the
constraints close, {\it i.e.,}
$$
\{\phi_i\,,\phi_j\} = f^k_{ij}\,\phi_k\,,
\eqn\first
$$
for some structure functions $f^k_{ij}$. In Yang-Mills theory it is
Gauss' law that plays the role of the constraints and the structure
functions are actually constants.

The idea behind Dirac's approach to the quantization of such a
constrained system is to first quantize on the extended system,
ignoring the constraints, then reduce to the physical states by
imposing the conditions implementing the constraints (\con)
at the quantum level:
$$
\widehat\phi_i\, \ket{\psi_\phy}=0\,.
\eqn\pstates
$$
As long as the first class nature (\first) of the constraints has been
preserved (and modulo ordering problems if the structure
{}functions are not constants),
condition (\pstates) is consistent and should result in a
reasonable set of physical states. In fact, it is expected that this
process will recover the same set of states arrived at by directly
quantizing on the true degrees of freedom.  So we can summarize
Dirac's approach to the quantization of a constrained system
by the (commutative) diagramme in Fig.\thinspace1.
This freedom of changing the ordering of quantizing and reducing is
directly testable in electrodynamics and, although one can find
toy models where curvature effects can spoil the commutivity of this
description [\Kuchar], more modern accounts which use ghost
variables [\MP] show that we are able in general to recover the expected
quantization on the reduced space from this type of constrained
quantization on the extended system.
But what is the expected quantization?

\topinsert
$$
\diagram{
\stmta&\harrow{quantization}{}&\stmtb \cr
\varrow{reduction}{}&&
\varrow{}{reduction} \cr
\stmtc&\harrow{}{quantization}&\stmte\cr
}
$$
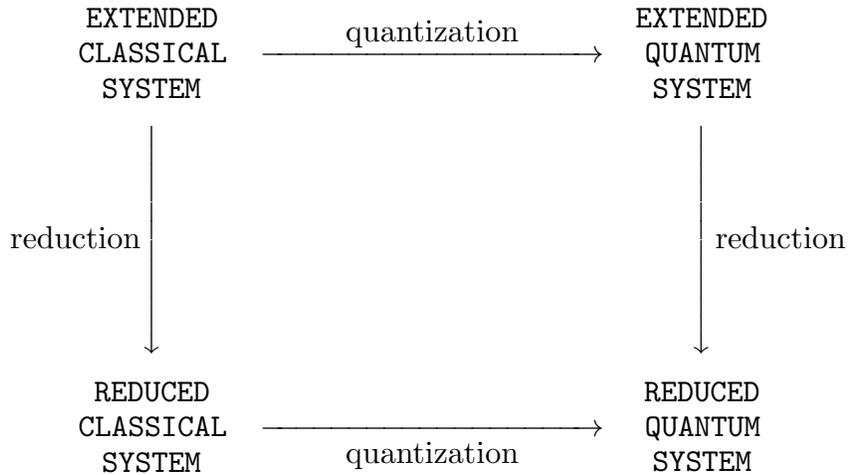
\centerline{\bf Fig.\thinspace1: Traditional point of view on
reduction/quantization}
\vskip 3mm
\endinsert

Our expectations for what is the correct quantization on the reduced
system is based, to a large extent, on our experience with
quantizing simple quantum mechanical systems such as the harmonic
oscillator or the Hydrogen atom --- systems which have very simple
configuration spaces $Q$ of the form $\Real^n$, for some $n$.
{}For electrodynamics the physical configuration space is essentially
a linear space, so we are quite confident in our intuition for what
is the expected quantum theory. But in Yang-Mills theory the
physical configuration space $Q$ is a very complicated space [\Singer],
having features more in common with the coset spaces of the form
G/H, for some groups G and H, rather than $\Real^n$, for any $n$.

Quantum theory on a coset space G/H is much richer than
that on $\Real^n$. Indeed, it was shown by Mackey [\Mackey] that
there are {\it many different quantizations} possible on G/H,
labelled by the irreducible unitary representations of the group H.
After Mackey's work of more than a quarter of a century ago, there
appeared a number of other approaches to the quantization on the coset
G/H (or, more generally, to the quantization of
a classical system whose configuration space $Q$ is more complicated
than $\Real^n$).  Among them are geometric quantization
[\Woodhouse], the canonical group approach [\Isham] and a ${\rm
C}^*$-algebric reformulation of Mackey's analysis [\LandsmanA, \LL],
all of which emphasize different structures and do
not appear to be equivalent.
Nonetheless, for the coset space G/H that we
are considering, they all conclude that
inequivalent quantizations do exist, and that these
different quantum sectors
always include those
described by Mackey; in this
sense we may regard Mackey's approach as the most fundamental one.
Recently, a significant advance in
understanding the dynamical consequences of these inequivalent
quantizations was taken by Landsman and Linden [\LL], who showed that
the different quantum sectors come equipped with a specific type of
Yang-Mills field, called the H-connection.

Most of these approaches, including Mackey's,
quantize directly on G/H by generalizing
the canonical commutation relations,
and share a
common feature that they deal with {\it vector-valued} wave
{}functions, rather than the usual scalar-valued ones.
As such they suffer severely from techincal difficulties, which
hamper the extention to field theory that we are aiming at.
In [\LL] an attempt was made at directly developing a path-integral
account of the various quantum sectors by starting with the
vector-valued wave functions.
Again, the problem is that, due to the vector-valued
nature of the states, the path-integral must always be path-ordered,
resulting in
a \lq discrete' path-integral leading to
a transition {\it matrix} rather than an amplitude.
Consequently,
one cannot recover the usual expression for the path-integral, that
is, as a summation over \lq continuous' paths weighted by
the exponential of a classical action.
That Mackey's original formulation might not be the most natural can
also be argued from the fact that the role of the H-connection
only became apparent {\it twenty two}
years after Mackey's book on the
subject was published!

The above discussions represent the dual themes of this paper: how
should Dirac's approach be extended to take into
account the possibility that there are many possible quantizations
on the reduced system, and what is the most natural (useful) formulation
of Mackey's analysis of the quantum theories on G/H? What we shall
{}find is that these two questions are, in fact, equivalent. That is,
in this paper we reformulate
Mackey's description of quantization
so that the use of vector-valued wave functions can be avoided,
thus allowing for the construction of a conventional continuous
path-integral description of the different quantum sectors.
This reformulation will be seen to
simplify the quantization significantly,
and clarify many of the
difficult aspects of Mackey's account of quantizing on G/H.
{}For instance, after our reformulation
the emergence of the H-connection and the gauge
structure associated with it ---
which we feel is central to
the physical understanding of the different quantum sectors and to the
extention to field theory ---
can be recognized almost immediately.  Furthermore,
the vector-valuedness that caused the trouble is
converted to a set of observables, called
\lq generalized spin', that have an effective classical
counterpart.
{}For our reformulation we shall adopt a constrained point of view,
that is, we shall view the system on G/H as a reduction of a larger,
albeit simpler to quantize,
system. We will then find that, in order to recover Mackey's
description of the possible quantizations, a generalized version
of  Dirac's approach is needed.
To our pleasant surprize, the generalization we need turns out to be
quite simple: one just replaces the physical state
conditions (\pstates) with
$$
\widehat\phi_i\, \ket{\psi_\phy} = K_i\, \ket{\psi_\phy}\,,
\eqn\npstates
$$
where the $K_i$ are constants (integers) characterizing the
irreducible unitary representations of the group H.

The conditions (\npstates) show that
upon quantization the constraints
become anomalous ({\it i.e.}, some of them become second class)
and should be treated as such.
This result obtained for
$Q \simeq {\rm G/H}$
leads us to the following conjecture for the quantization
on a generic configuration space $Q$:
one should, in general, expect many inequivalent quantizations
on $Q$ if it is \lq non-trivial'; these quantizations can be
obtained by the generalized Dirac approach, that is, {\it first
view  the system as a reduction from  a \lq trivial' one,
whose quantization is simple, and then reduce it
at the quantum level, taking into account the allowed
anomalous nature of the constraints.}
Accordingly, the previous diagramme, Fig.\thinspace1, is no
longer appropriate for this constrained approach to
quantizating with  a generic
reduced configuration space $Q$, and needs to be replaced by
the diagramme in Fig.\thinspace2.  Of course, for this
generalization to
work a number of problems
should be solved; for instance, we must find a suitable
extended configuration space which is trivial in the above sense,
and specify what the anomalous $K_i$ are.
{}For the case $Q \simeq {\rm
G/H}$ our natural choice for the extended configuration space
will be the group G, and it will be shown in this paper that
the $K_i$ are actually determined correctly from a
consistency of the path-integral.
Interestingly,
this generalized Dirac approach
can be applied to Yang-Mills theory (where again there is a natural
choice for the extended configuration space) using the conditions
(\npstates) with field-dependent $K_i$, which yields
additional topological terms in the Yang-Mills Lagrangian
as a consequence of the inequivalent quantizations [\MT].
We also mention that this change in the nature of the constraints
is familiar when quantizing anomalous gauge theories (see [\Fadshat]
and also the discussion in [\Jackiw]),
which indicates a possible new response to gauge anomalies based on
the possibility of exploiting the existence of inequivalent quantizations.
In this paper, however,
to establish a firm
ground for those extentions, and to examine the physical
implications of inequivalent quantizations as clearly as possible,
we shall confine our arena to the coset spaces $Q \simeq {\rm G/H}$,
which is also worth investigating in its own right.

\topinsert
$$
\diagram{
\stmta&\harrow{quantization}{}&\stmtb \cr
\varrow{reduction}{}&&\varrows{}{many reductions} \cr
\stmtc&\harrows{}{many quantizations}&\stmtd\cr
}
$$
\centerline{\bf Fig.\thinspace2: Generalized version of Dirac's
approach}
\vskip 3mm
\endinsert

The plan of this paper is as follows. After this introduction, in
Sect.\thinspace2, Mackey's quantization on G/H will be described
along with some discussion of how it relates to other direct
approaches and
some of the problems with its usual description.
The aim of this somewhat lengthy review part is to provide
a self-contained account of Mackey's formulation,
which is later used to prove the equivalence to our (generalized
Dirac) approach.
In Sect.\thinspace3 we illustrate the idea of our
reformulation of Mackey's inequivalent quantization by a simple
example, where the emergence of spin
is explained from particle dynamics on $\Real^3$.
This will end up with a derivation of the path-integral description
of spin, proposed by Nielsen and Rohrlich [\NR],
{}from Mackey's account of quantization applied to the coset space
$\rm {\widetilde{E(3)}/SU(2)} \simeq \Real^3$.
After this illustration, in Sect.\thinspace4 we
present for the general coset
space a precise definition
of our generalization of Dirac's approach, and prove
its equivalence with Mackey's account
of quantizing on such coset spaces.
In Sect.\thinspace5 we
will develop the path-integral version of our approach presented in
Sect.\thinspace4, in which the effects of inequivalent quantizations
--- the H-connection and generalized spin --- become transparent.
This section can almost be read
independently of what has gone before (although Sect.\thinspace4.1
is needed for the notation),
and we would suggest to the reader that if one is only interested in
the final results of our analysis and the applications, then go
directly to Sect.\thinspace5 to see how simple our methods are.
Sect.\thinspace6 is devoted to a study of
the physical implications of inequivalent quantizations by examining
in detail the case $S^n \simeq {\rm SO(n+1)/SO(n)}$ for $n = 2$, 3, 4.
The physical interpretation of
spin will be given for $S^3$ and $S^4$, and,
in particular on
$S^4$, it will be shown that chiral spin degrees
of freedom couple to a BPST instanton (and anti-instanton),
which is the H-connection in this case.
{}Finally, in Sect.\thinspace7, we shall
conclude this analysis with some speculations on the possible
ramifications of the ideas presented in this paper. There are also
{}four appendices containing some technical results used in the main
text, the most important of which is Appendix\thinspace A which,
among other things, gives our notational conventions for dealing
with the many spaces needed in this account of quantization.

\vfill\eject

\secno=2 \meqno=1

\noindent
{\bf 2. Direct approaches to quantizing on coset spaces}
\medskip
\noindent
In this section the quantum mechanics of a particle moving on a
coset space G/H is reviewed.  Among the many approaches that quantize
directly on G/H, we
discuss primarily Mackey's, which is
the first
and the most fundamental one in the sense mentioned earlier.
Special emphasis will be placed on
the emergence of inequivalent quantizations, labelled by the
irreducible unitary representations of the group H, and the
effect this has on the dynamics in the different sectors.
Later, we shall briefly discuss other approaches together with
problems that arise with the
applications and extensions of these results.

\bigskip
\noindent
{\it 2.1. Mackey's approach: the system of imprimitivity}
\medskip

A spinless particle moving on the flat configuration space $\Real^3$
provides the most familiar example of the quantization process. The
basic observables of this system are those of  position
$\hat q^\a$ and momentum $\hat p_\a$ $(\a=1,2,3)$. These
self-adjoint operators satisfy the canonical commutation relations,
$$
[\hat q^\a,\, \hat q^\b] = 0\,,
\qquad
[\hat p_\a\,,\, \hat p_\b]= 0\,
\qquad\hbox{and}\qquad
[\hat q^\a,\, \hat p_\b]=i\d^\a_\b\,,
\eqn\ccr
$$
which reflects the classical Poisson bracket relations between
position and momentum.

To discuss the representations of these commutator relations,
and to address the question of their uniqueness, it is best to work
with the bounded unitary operators
$U(a) := \exp(-ia{\cdot}\hat p)$ and
$V(b) := \exp(ib{\cdot}\hat q)$ where $a,b \in \Real^3$,
in terms of which the
canonical commutation relations (2.1) become the Weyl relations:
$$
U(a+b) =  U(a)\, U(b)\,,
\qquad
V(a+b) =  V(a)\, V(b)
\eqn\WeylA
$$
and
$$
U(a)\, V(b)\,  U^{-1}(a) =
e^{-ia\cdot b} V(b)\,.
\eqn\Weyl
$$
A particular representation of these is furnished by the Schr\"odinger
representation in which the quantum states are identified with the
space $L^2(\Real^3)$ of wave functions $\psi(q)$ on the configuration
space, and the  unitary operators $U(a)$ and $V(b)$
become
$$
(U(a)\psi)(q) = \psi(q-a)\,,
\qquad
(V(b)\psi)(q) = e^{ib\cdot q}\psi(q)\,.
\eqn\Schro
$$
In this representation
the position observables $\hat q^\a$ acts on these states as the
multiplication operators, while the momentum observables $\hat p_\a$
become the derivative operators $-i\partial/\partial q^\a$.
The Stone--von Neumann theorem then states that all other irreducible
unitary representations of the Weyl relations, (\WeylA) and
(\Weyl), are unitarily
equivalent to the Schr\"odinger representation~(\Schro).

There are, however, various unsatisfactory aspects to this description of
quantization --- problems which become serious  when we attempt to
generalize this construction to more interesting systems. The most
obvious problem is that, for a general system whose configuration
space is \lq non-trivial\rq, we do not expect
there to
be globally well defined observables satisfying the canonical
commutation relations (\ccr). Another problem is that we seem
to be imposing on the quantum theory the restriction that the
observables be constructed solely out of the position and momentum
variables, and thereby excluding the possibility of spin
(or even other, more exotic, observables). What is surprising,
then, is
that both of these problems can
be resolved, for a wide class of theories, by simply reinterpreting
the Weyl relations, (\WeylA) and (\Weyl).

To understand this reinterpretation let us recall that in (Dirac's
account of) the Schr\"o\-dinger representation the abstract
quantum states $\ket\psi$ are expanded in terms of the position
eigenstates $\ket q$: $\hat q^\a\ket q=q^\a\ket q$, and the wave
functions are identified with the components of the state in this
position basis,  $\psi(q)=\<q\ket\psi$. The translation
group ${\rm G}=\Real^3$ acts transitively on the configuration space
$\Real^3$
with an action $q\mapsto gq:=q+g$ (we use here $g$ instead of $a$ to
stress its group property). The unitary operator $U(g)$
is then defined by
$$
U(g)\ket q=\ket{gq}\,.\eqn\uop
$$
Acting on the wave functions this becomes
$$
(U(g)\psi)(q)=\bra q
U(g)\ket\psi=\<{g^{-1}q}\ket\psi=\psi(g^{-1}q)\,.\eqn\no
$$
If $F(q)$ is any bounded function on the configuration space $\Real^3$
then the spectral theorem asserts that $\widehat F(q) := F(\hat q)$,
the operator
corresponding to it, can be written as
$$
\widehat F(q) = \int_{\Real^3}d^3q\, F(q)\ket q\!\bra q\,,
\eqn\specthm
$$
where the formal expression, $d^3q\ket q\!\bra q$, is to be identified
with the spectral measure associated with the self-adjoint
operator $\hat q$.
Using (\uop) and the translation invariance of the measure
on $\Real^3$, we see that
$$
U(g)\, \widehat F(q) \, U^{-1}(g)= \widehat
{}F(g^{-1}q)\,.
\eqn\imprim
$$
A unitary representation $U(g)$
of the Lie group ${\rm G}=\Real^3$ satisfying
the relation (\imprim), where
$F$ is any bounded function on $\Real^3$, was called by
Mackey\note{Mackey's formulation of a system of imprimitivity was
slightly different from the (simplified) one given here;
it was not based on the bounded functions but on the projection valued
measure $E_\triangle$, where $\triangle$ is a Borel subset of the
configuration space, and a system of imprimitivity referred to
the pair $(U(g), E_\triangle)$.
In terms of these the imprimitivity relation
(\imprim) takes the form
$ U(g)E_\triangle  U^{-1}(g)=E_{g\triangle}$,
where $g\triangle$ is the translated Borel set.
{}For the systems we
consider in this paper, this description in terms of projection valued
measures can be seen to be equivalent to the one  based on the bounded
{}functions.
Indeed the projection valued measure $E_\triangle$ can be
recovered from (\specthm) since
$E_\triangle\equiv\int
d^3q\, \chi_{_{\scriptstyle\triangle}}(q)\ket q\!\bra q$, where
$\chi_{_{\scriptstyle\triangle}}$
is the characteristic functions for the Borel subset
$\triangle$  of the configuration space.}
a system of imprimitivity based on~$\Real^3$
[\Mackey] (see also [\Varadarajan]).
{}For this abelian group acting
transitively on the configuration space Mackey showed that
there is a unique
representation of this system of imprimitivity, which is just
the Schr\"odinger
representation presented above.
This reformulation of the
Stone${-}$von Neumann theorem by Mackey has allowed us to replace
the specific bounded operator $V$ which
satisfies the relation
(\Weyl) (the existence of which is
closely related to the geometry of $\Real^3$) by an {\it arbitary\/}
bounded function $F$ whose operator $\widehat F$ satisfies
the imprimitivity relation (\imprim).

This construction has an immediate generalization.
Consider a configuration space on which a group
G acts transitively\note{In order to
continue using bounded functions in place of projection valued
measure in the imprimitivity relation we need to assume that the
action of G on the coset space G/H is continuous and that G/H
is locally compact Hausdorff and satisfies the second axiom of
countability (see lemma 9.1 in Ref.\thinspace\Varadarajan). This
will be the case
if G compact, or locally compact and abelian, or a (semidirect)
product of such groups.}.
If a closed subgroup ${\rm H} \subset {\rm G}$
is the isotropy group of the
action then we can identify the configuration space with the
coset space G/H.
We call a unitary representation $U(g)$ of G
a {\it system of imprimitivity based on} G/H
if it satisfies the relation (\imprim) for
any bounded function $F(q)$ on G/H.

For example, the translation group $\Real$ does not act
transitively on the configuration space $\Real^+$  (the positive real
line), rather an affine action is needed: $q\rightarrow\lambda q$, for
$\lambda\in\Real^+$. Thus, although
operators satisfying the Weyl relations (\WeylA) and (\Weyl)
cannot be found for $\Real^+$, a system of imprimitivity
can be constructed using this affine action~[\Isham].
A less trivial example follows from the simple
observation that the Euclidean group
E(3)$\,=\,$SO(3)$\,\ltimes\Real^3$ also acts transitively on $\Real^3$;
so if we wish to study quantum
systems invariant not just  under translations but under this
larger group then we should start with
the system of imprimitivity based on $\Real^3$, now viewed as the coset
space E(3)/SO(3). This example serves to highlight the basic fact
that there is no  unique system of imprimitivity to be assigned
to a system --- reflecting the fact that the identification of a coset
space as G/H is
{}far from unique.  Thus the choice of a specific system of imprimitivity
must be guided by additional requirements such as symmetry or
dynamics.

Now that the imprimitivity relation (\imprim) is taken to be the basis
upon which the quantum theory of a system with configuration space
G/H is constructed, we would like to know as much as
we can about the irreducible, unitary representations of this
relation. In what follows we shall show that, in
contrast to the Stone--von Neumann theorem, there are now many
inequivalent representations of (\imprim) possible, labelled by the
irreducible unitary representations of H, and hence many
inequivalent quantizations for a system based on G/H.
In order to do this we first need to recall some properties of coset spaces.

To fix the geometry of the coset space (and our notation), the group
G is taken to act on the left, hence the coset space G/H
is identified with the left cosets $\{g{\rm H}\, \vert \, g\in
{\rm G}\}$.
One can regard G as the principal fibre bundle G(G/H, H) with the
base space G/H , the fibre H and
the projection from G to G/H being $\pro(g):=g$H.
Note that for the left action of G we have
$\pro(g_1g_2)=g_1\pro(g_2)$,
while for the right action of H we have $\pro(gh)=\pro(g)$.
A section $\s(q)$ is a map $\s:{\rm G/H}\mapsto{\rm G}$ such that
$\pro(\s(q))=q$. We just require that the map $\s$ be measurable,
allowing $\s$ to be defined only locally as a continuous section.
Given a section $\s$ then $g\s(q)$ is an element in the fibre
above $gq$, and hence we have $h_\s(g,q) \in {\rm H}$ such that
$$
g\s(q)=\s(gq)h_\s(g,q)\,.
\eqn\hsigma
$$
If $\s'$ is another section then $\s'(q) = \s(q) \tilde h(q)$ for
some $\tilde h(q)\in {\rm H}$.
It then follows that under a change of section we have
$$
h_{\s'}(g,q)= \tilde h^{-1}(gq) h_\s(g,q) \tilde h(q)\,.
\eqn\csection
$$
When it is clear, or not too important, which section is involved we
shall simply write
$h_\s(g,q)$ as $h(g,q)$.
As well as this behaviour (\csection)
under the change of section,
$h(g,q)$ also
satisfies the condition $h(1,q)=1$ and the cocycle condition
$$
h(g_1g_2,q)=h(g_1,g_2q)h(g_2,q)\,.\eqn\cocycle
$$

The coset space G/H has a naturally defined G-invariant metric
induced from the Killing metric on G (see Appendix C
{}for full
definitions). Thus there is a
G-invariant measure $Dq$ on G/H which can be characterized by
$$
\int_{\rm G} Dg\, F(g)
= \int_{{\rm G/H}}\!\! Dq
  \int_{\rm H}  Dh\, F(\s(q)h)\,,
\eqn\measure
$$
where $F$ is an arbitrary function on G and $Dg$ ($Dh$) is the Haar
measure on G (H).

Given an irreducible unitary representation $\chi$ of H on a
Hilbert space $\H_\chi$,
we can construct the Hilbert
space  $\H$ over G/H consisting of $\H_\chi$-valued state vectors
on G/H, {\it i.e.,}
$$
\H \simeq L^2({\rm G/H},\H_\chi)\,.
\eqn\no
$$
Given such a ket
$\ket{\psi}\in\H$, we can find its components in the standard way;
we use the position basis $\ket{q}$ for G/H and the finite
dimensional
orthonormal basis $\ket{\chi,\mu}$ for $\H_\chi$
(for notations, see Appendix A)
and define the {\it vector-valued} wave functions to be
$$
\psi_\mu(q) = (\bra{q}\otimes\bra{\chi,\mu})\ket{\psi}
=\langle{q,\chi,\mu}\ket{\psi}\,.
\eqn\vvwf
$$
The inner-product on these states is constructed from the pairing on
$\H_\chi$ and the G-invariant measure $Dq$ on G/H:
$$
\<\phi\ket\psi=\int_{\rm
G/H}\!Dq\,\<\phi(q)\ket{\psi(q)}_{_{\scriptstyle\H_\chi}}\,.\eqn\no
$$
On these vector-valued wave functions we can define the unitary
{\it induced representation} $U(g)$ of (the left action of)
G by
$$
(U(g)\psi)_\mu(q)= \sum_\nu \pi^\chi_{\mu\nu}
(h_\s(g,g^{-1}q))\,\psi_\nu(g^{-1}q)\,,
\eqn\induced
$$
where the matrix elements of the unitary operator $\pi^\chi(h)$,
implementing the irreducible representation $\chi$, are
$$
\pi^\chi_{\mu\nu}(h):=\bra{\chi,\mu}\pi^\chi(h)\ket{\chi,\nu}\,,
\eqn\matel
$$
and a choice of section has been made. The fact that
this is a representation follows from the cocycle condition
(\cocycle).

In the representation $U(g)$ in (\induced)
we have suppressed reference to the
choice of section $\s$ since a different choice of section leads to a
unitarily equivalent representation.
To see this we write $U(g)=U_\s(g)$,
to expose the $\s$-dependence explicitly, and consider the unitary mapping
$T:\H\to\H$ given by
$$
(T\psi)_\mu(q)= \sum_\nu
\pi^\chi_{\mu\nu}(\tilde h(q))\,\psi_\nu(q)\,,
\eqn\repspin
$$
where $\tilde h(q)$ is the element of H relating the section $\s$ to the
section $\s'$ as in (\csection).
Then one can readily confirm that $T$
intertwines the two representations,
$$
(U_\s(g)T\psi)_\mu(q)
=(TU_{\s'}(g)\psi)_\mu(q)\,,
\eqn\no
$$
hence they are unitarily equivalent.

It is now clear that this unitary representation of G --- induced
from the irreducible representation $\chi$ of H --- provides us
with a representation of the imprimitivity relation (\imprim) since, for
any bounded function $F(q)$ on G/H,
$$\eqalign{
\bigl(U(g)\,\widehat F(q)\, U^{-1}(g)\psi\bigr)_\mu(q)
&= \sum_\nu \pi^\chi_{\mu\nu} (h(g,g^{-1}q))\,
\bigl(\widehat F(q)\, U^{-1}(g)\psi\bigr)_\nu(g^{-1}q)\cr
&= \sum_\nu \pi^\chi_{\mu\nu}
(h(g,g^{-1}q))\,F(g^{-1}q)\bigl(U^{-1}(g)\psi\bigr)_\nu(g^{-1}q)\cr
&= \sum_{\nu,\lambda} \pi^\chi_{\mu\nu}
(h(g,g^{-1}q))\,F(g^{-1}q)\,\pi^\chi_{\nu\lambda}(h(g^{-1},q))\,
\psi_\lambda(q)\cr
&= \sum_\lambda \pi^\chi_{\mu\lambda}(h(g,g^{-1}q)h(g^{-1},q))
\, F(g^{-1}q)\,\psi_\lambda(q)\cr
&= F(g^{-1}q)\, \psi_\mu(q)\cr
&= \bigl(\widehat F(g^{-1}q)\psi\bigr)_\mu(q)\,.
}\eqn\no
$$
The imprimitivity
theorem due to Mackey then states that these induced
representations provide {\it all\/} the distinct irreducible
representations of the imprimitivity relation (\imprim).
Thus we see that, in
contrast to the uniqueness of the Schr\"odinger
representation based on the Weyl relations (\WeylA), (\Weyl),
{\it there are many inequivalent quantizations possible}
when the configuration space is identified with the coset space
G/H, and that these inequivalent quantizations are labelled by
the irreducible unitary representations of the group H.

Returning to the example of $\Real^3$, which we identified with
the coset space E(3)/SO(3), we see that there are
now inequivalent quantizations labelled by
the irreducible unitary representations of SO(3).
Alternatively,
to avoid the need to consider projective
representations of (\imprim), we may work with the universal
covering group
$\widetilde{{\rm E(3)}}
\equiv\,$SU(2)$\,\ltimes\Real^3$, thus identifying $\Real^3$ with the
coset $\widetilde{{\rm E(3)}}$/SU(2). Hence it is
the positive (half-) integers labelling the irreducible unitary
representations of
SU(2) that characterize the possible quantizations.
Reverting to more familiar notation, we know
that, for each half-integer $j$, there is an irreducible
representation $\pi^j=D^j$ of SU(2) on the Hilbert space
$\H_j\simeq\C^{2j+1}$. Thus the  quantization of this
system of imprimitivity corresponding to the irreducible
representation of SU(2) labelled by  $j$ is defined on
the Hilbert space
$\H \simeq L^2(\Real^3,\C^{2j+1})$.
Denoting $g\in\widetilde{{\rm E(3)}}$ by
$g=( h, a)$ where $h \in {\rm SU(2)}$, $a \in \Real^3$,
and choosing the section $\s(q)=(1,q)$,
we see from (\induced) that the representation of
$\widetilde{{\rm E(3)}}$ in
this quantization  is given by
$$
\bigl(U(h,a)\psi\bigr)_m(q)= \sum_{n = -j}^j
D^j_{mn}(h)\,\psi_n(h^{-1}
(q-a))\,.\eqn\spin
$$
The Schr\"odinger representation (\Schro)
simply corresponds to the case $j=0$,
that is, the
trivial representation of SU(2).

The unitary representation (\induced) of G that arises
in the system of imprimitivity
allows us to define left-momentum operators
$\widehat\L_m$
($m=1, \dots, {\rm dim\,G}$) by
$$
(\widehat\L_m\psi)_\mu(q) := \frac{d\ }{dt}
\Bigl({U}(e^{t T_m}) \psi \Bigr)_\mu
(q)_{\displaystyle|_{\scriptstyle t=0}}\,,
\eqn\inducealgebra
$$
where $\{ T_m \}$ is a basis of the Lie algebra $\g$ of G
and $e^{t T_m}$ is the exponential mapping on G.
This representation of the Lie algebra $\g$ can be extended to the
enveloping algebra, allowing us to construct observables polynomial in
the momentum.
We note that
when this identification of momentum is applied to the
representation (\spin) of $\widetilde{{\rm E(3)}}$
we recover, in addition to the observables corresponding to the
generators of translations $\hat p_\a$, new basic
observables $\widehat S_i$ ($i=1,2,3$) satisfying
$[\widehat S_i\,,\widehat S_j]=i\e_{ijk}\widehat S_k$.
Clearly, $\widehat S_i$ can
be regarded as
spin in the quantum theory; in fact, this account of spin based on
inequivalent quantizations was
one of the prime motives for Mackey's original analysis [\Mackey].

\bigskip\noindent
{\it 2.2. Dynamics and the {\rm H}-connection}
\medskip

Having addressed the kinematical aspects associated
with the
quantization of a system whose configuration
space $Q$ is G/H, we now wish to discuss how the dynamics is described
in the various inequivalent quantizations.
{}For simplicity we shall consider the Hamiltonian
leading to a free particle moving on the coset G/H.
But before discussing a generic coset G/H, let us recall that
in the ${\rm G/H = \widetilde{E(3)}/SU(2)} \simeq \Real^3$ case
the Hamiltonian for a free particle is identified with the operator
$\widehat H = \frac12\hat p_\a\hat p_\a$, and this
can be interpreted as the representation furnished by
(\inducealgebra) for the
quadratic Casimir operator for
${\rm \widetilde{E(3)}}$ restricted
to its abelian, normal subgroup $\Real^3$.  Similarly,
{}for a generic G/H with
compact G, the Hamiltonian $\widehat H$
may be identified with the quadratic
Casimir of the group G; indeed, it is known that the classical
Hamiltonian $H$, that corresponds to that $\widehat H$, generates
{}free (geodesic) motion on G/H.  Thus, for such systems, the
Hamiltonian is given by
$$
\widehat{H} = \tfrac12 \eta^{mn} \widehat{\L}_m \widehat{\L}_n\,,
\eqn\qham
$$
where $\eta^{mn}$
is the inverse of $\eta_{mn} := \Tr (T_m T_n)$.
In passing we note that
considering a Hamiltonian for non-free particle by adding
potential energy terms to (\qham)
does not siginificantly alter the conclusions we shall reach.

Through an analysis of the form of the Hamiltonian (\qham), Landsman
and Linden [\LL] (see also [\Koch, \Landsman, \Ohnuki, \Tanimura])
made the important observation
that {\it there is a \lq fictitious'
connection that couples to the
particle in the non-trivial quantum sectors} among the inequivalent
quantizations.
More precisely, they showed
that in the Hamiltonian
(\qham) one can isolate the quadratic
Casimir for the group H which becomes constant on
a irreducible representation
$\chi$.  Then, using some local set of coordinates on G/H,
the Hamiltonian can be written (up to the constant)
in the gauge covariant form,
$$
\widehat
H=-\frac12g^{\a\b}(\nabla_\a+A_\a)(\partial_\b+A_\b)\,,
\eqn\qhamcon
$$
where $\a=1,\dots,\, {\rm dim\,(G/H)}$, $\nabla_\a$ is the covariant
derivative construced
out of the natural G-invariant metric $g_{\a\b}$ on
G/H, and the connection $A_\a$ is the canonical H-connection.
Note that in a non-trivial sector (corresponding to a non-trivial
representation $\chi$)
the Hamiltonian $\widehat H$
is a matrix-valued ($\pi^\chi(\h)$-valued) vector field on G/H,
while in the trivial sector it reduces to
the Laplace-Beltrami operator,
$\widehat H = -\frac12 \triangle_{\rm LB} =
-\frac12 g^{\a\b}\nabla_\a\partial_\b$, as required.
The precise definition, and properties, of the
H-connection will be
given in detail later, but for now let us mention
a simple example where the connection becomes a familiar object.
Consider a particle moving freely on the two-sphere $S^2$.
The configuration space can be identified with the coset space
$S^2 \simeq {\rm SO(3)/SO(2)}$,
thus the quantum theories will be characterized by an
integer $n$ labelling the irreducible unitary representations
of SO(2). The resulting SO(2) connection on $S^2$ can be identified
with that of a Dirac monopole centred at the origin of the two-sphere.
Thus, for this specific example, we can interpret the
different possible quantum theories as those that describe the
particle coupled to a Dirac monopole of magnetic charge~$n$
(for more details, see Sect.\thinspace6.3).
\goodbreak
\bigskip
\noindent
{\it 2.3. Other approaches and problems}
\medskip

Mackey's account of quantization is, albeit the first, just one of
several approaches to the quantization of a classical system whose
configuration space is more complicated than $\Real^n$.
As we mentioned in the Introduction, these other approaches
include
geometric quantization
[\Woodhouse], the canonical group approach [\Isham] and a ${\rm
C}^*$-algebra approach [\LandsmanA, \LL].
These have their own virtues but
{}for the coset spaces G/H
they all reproduce the inequivalent quantizations
described by Mackey through his system of imprimitivity.
Take for example a particle
moving freely on the two-sphere $S^2$.
The methods of geometric quantization, which gives
an account of the transition from a classical
to a quantum system based on the (symplectic) geometry of the
phase space, predicts many
quantizations
labelled by the cohomology group $H^2(S^2,\Z)\simeq\Z$
classifying the possible prequantum line bundles on the phase space
$T^*S^2$ (the cotangent bundle of $S^2$).
If we follow the canonical group approach
[\Isham] and exploit the fact that the phase space $T^*S^2$ is,
in fact, a homogeneous symplectic manifold upon
which the (canonical) group
E(3) acts transitively and symplectically, then,
the possible quantizations will
correspond to the irreducible representations of the canonical group
E(3) which, by
Mackey's theory of induced representations applied to such a
semi-direct product group [\Mackey], are labelled by the
irreducible representations of the isotropy (little)
group SO(2).
The connection between
geometric quantization, the canonical group approach and Mackey's
approach becomes even
clearer when we recall that the irreducible representations of E(3)
are geometrically characterized by the coadjoint orbits of E(3).
Generically, these
orbits are precisely the manifold $T^*S^2$, but with a non-standard
symplectic form corresponding to that describing the minimal
coupling of a particle with a
Dirac monopole (see page 14 in [\Perelomov]). So we see that, at least
{}for the simple
model of a particle on $S^2$, there seems to be a consensus
in the literature that many inequivalent quantizations are possible,
and that these different quantum sectors come equipped with a
topological abelian gauge
{}field.  For the more general coset space we can again argue that
Mackey's account of the inequivalent quantizations will at least be
contained in the other approaches since the phase space $T^*$(G/H) are
also homogeneous symplectic manifolds [\Isham]. It is, however,
important to note that the representation theory of the appropriate
canonical group, or, equivalently, geometric quantization applied to
the various coadjoint orbits,
will not always result in the form of
vector-valued wave functions on the original coset
space, thus we must always keep it in mind that it is possible for
more to happen in the process of quantizing a classical system than
Mackey would suggest. (For more details on this point
we refer to [\Isham, \LL].)

The approaches to quantization discussed above all take as basic the
need to generalize the canonical commutation relations (\ccr), and
thus can be characterized as being, at heart, \lq operator
descriptions'.  As such they lack the physical appeal of the
path-integral approach to quantization and, hence, suffer from all the
problems inherent in such operator formalisms when one tries
to generalize these quantum mechanical results to field theory.
In particular,
the vector-valued
nature of the wave functions forces
the directly developed path-integral [\LL] to be unattractive; it is
a type of path-ordered, discrete path-integral which
leads to a transition matrix rather than an amplitude.
In fact, such a
problem was encountered earlier when spin was
incorporated into the path-integral [\Schulman],
where, in order to avoid the use of
the discrete path-integral,
an attempt was made to develop a continuous path-integral for spin.
Thus it will be natural to illustrate the idea of our reformulation
of Mackey's approach by showing how spin can emerge
in the context of quantizing on a coset, and how the
continuous path-integral can be developed which reproduces all the
results that the operator descriptions provide.  This will be the subject
of the next section.

\vfil\eject

\secno=3 \meqno=1

\noindent
{\bf 3. Spin without vector-valued wave functions}
\medskip
\noindent
In this section an account of spin will be presented that views the
coset space $\Real^3\simeq \widetilde{{\rm E(3)}}$/SU(2) as a constrained
system on $\widetilde{{\rm E(3)}}$. This will be seen to be equivalent
to Mackey's description in
terms of a system of imprimitivity, but avoids the use of vector-valued
wave functions. The path-integral version of the construction
will be seen to recover the continuous spin description of Nielsen and
Rohrlich [\NR], although the argument for the quantization of spin will be
different. The methods developed in this section will form the
basis for our generalization of Dirac's approach to
include the possibility of inequivalent quantizations.

\bigskip
\noindent
{\it 3.1. An operator description of spin}
\medskip

We have seen that through Mackey's analysis, spin enters into
the quantum theory once the
configuration space $Q = \Real^3$ is identified with the coset space
$\widetilde{{\rm E(3)}}$/SU(2). The resulting quantum
states become vector-valued wave functions, and new basic
observables $\widehat S_i$ emerge which are identified with the spin.
In this
account spin arises only in the quantum theory --- no classical
counterpart exists.  However, from a path-integral point of view, it
is desirable to avoid using vector-valued wave functions and,
instead, introduce an (effective) classical system whose
quantization leads to the observables $\widehat S_i$ and the finite
dimensional Hilbert spaces $\H_j\simeq\C^{2j+1}$.

The first attempt to develop such a classical account of spin was
by Schulman [\Schulman]. Focusing purely on the spin degrees of
{}freedom, we see that
the group SU(2) can itself be taken to be a configuration space
of a system that has momentum variables obeying the su(2) algebra, and thus a
good candidate for spin. Applying Mackey's quantization procedure to
this system (where now ${\rm G/H = SU(2)}/\{1\}$) we get a single
quantization in which
the state space is identified with the wave functions on SU(2):
$\H\simeq L^2$(SU(2)).
The unitary representation (\induced) of SU(2) is
now the left-regular representation
$$
(U_{\rm L}
(\tilde g)\psi)(g)=\psi(\tilde{g}^{-1} g)\,,\eqn\lregular
$$
where both $\tilde g$ and $g$ are elements of SU(2).
The generators $\widehat\L_i$ ($i=1,2,3$) of this left action
satisfy the su(2) commutator relations
$$
[\widehat\L_i\, ,\widehat\L_j]=i\e_{ijk}\widehat\L_k\,.\eqn\no
$$
Although this quantum theory proposed by Schulman has basic observables
satisfying the
commutator relations for spin, it is not a satisfactory account
since the state space is not finite dimensional. In fact {\it all\/}
irreducible representations of SU(2) (and hence all possible total
spins, not just a specific one) are contained
in the Hilbert space $L^2(SU(2))$.
This result follows at once from the
Peter-Weyl theorem applied to SU(2), which we now summarize.

The content of the Peter-Weyl theorem [\Barut] is that the matrix elements
$D^j_{mn}(g):=\bra{j,m}D^j(g)\ket{j,n}$, of
the irreducible representations $D^j(g)$ of SU(2), form an orthonormal
basis for the wave
{}functions on SU(2). Thus
any wave function
$\psi(g)\in L^2(\rm{SU(2)})$
can be written as
$$
\psi(g)=\sum_j \sum_{m = -j}^j \sum_{m' = -j}^j
              C^j_{mm'}D^{j*}_{mm'}(g)\,,
\eqn\basis
$$
where the matrix elements satisfy the orthonormal relations
$$
\int_{\rm SU(2)}\!\!Dg\,D^{j*}_{mm'}(g)D^j_{nn'}(g)=\frac1{2j+1}\d_{mn}
\d_{m'n'}\,,
\eqn\no
$$
and $Dg:=\prod_{i=1}^3(g^{-1}dg)^i$ is the Haar measure on SU(2).
Thus we see that the wave functions $\psi(g)$ do contain all
spins $j$ in general.  Incidentally, we note that
the use of $D^{j*}_{mm'}(g)$ in (\basis)
is forced on us by the requirement that
we wish to construct wave functions which transform correctly under
the left action of SU(2), namely, we wish to have the identities
$$\eqalign{
\widehat\L_3 \,D^{j*}_{mm'}(g)&=mD^{j*}_{mm'}(g)\,,\cr
\widehat\L_\pm\, D^{j*}_{mm'}(g)&=\sqrt{(j\mp m)(j\pm m+1)}
\,D^{j*}_{m\pm1,m'}(g)\,.
}\eqn\identleft
$$

To improve upon Schulman's analysis we must find some means of
isolating a finite dimensional subspace of these wave functions
which carries a single irreducible representation of SU(2). To
{}facilitate in this we recall that,
as well as the left action of SU(2) on itself, there is also a right
action which leads to the right-regular representation of SU(2),
$$
(U_{\rm R}
(\tilde g)\psi)(g):=\psi(g \tilde g)\,.\eqn\rregular
$$
The generators of this right action, which we denote by $\widehat\R_i$,
also satisfy the su(2) commutator relations
$$
[\widehat\R_i\,,\widehat\R_j]=i\e_{ijk}\widehat\R_k\,,\eqn\no
$$
and commute with the generators of the left action
$$
[\widehat\R_i\,,\widehat\L_j]=0\,.\eqn\lrcommute
$$
Analogously to (\identleft),
acting on the wave functions $D^{j*}_{mn}(g)$ we have the identities
$$\eqalign{
\widehat\R_3\, D^{j*}_{mm'}(g)&=m'D^{j*}_{mm'}(g)\,,\cr
\widehat\R_{\pm}\, D^{j*}_{mm'}(g)&=\sqrt{(j\pm m')(j\mp m'+1)}
\,D^{j*}_{m,m'\mp1}(g)\,.
}\eqn\rup
$$
{}From (\basis) and (\rup) we see that the wave functions on SU(2) which
satisfy the two conditions
$$
\widehat\R_3\psi(g)=j\psi(g)\qquad{\rm and}\qquad
\widehat\R_-\psi(g)=0\,,\eqn\conditions
$$
are of the form
$$
\psi(g)=\sum_{m=-j}^j C^j_{mj}D^{j*}_{mj}(g)\,.\eqn\spinstates
$$
Thus the solutions to (\conditions) span a finite dimensional
Hilbert space (of dimension $(2j+1)$) with basis vectors given by the
wave functions $\{D^{j*}_{mj}(g)\}$ ($m=-j,\dots,j$). Due to
(\lrcommute),
the operators $\widehat\L_i$ preserve these states and hence, on
this subspace,
provide an irreducible representation of SU(2) with highest weight
(total spin) $j$.
The generators of spin, $\widehat S_i$, are thus identified
with the restriction to these states of the generators
$\widehat \L_i$ of the
left-regular representation of SU(2).

The conclusion of this analysis is that the quantum description of
spin emerges from the system of SU(2) subject to the restrictions
(\conditions) on the states.
This operator description of spin uses wave functions on SU(2)
subject to the conditions (\conditions), hence a traditional path-integral
formulation should be possible.

\bigskip
\noindent
{\it 3.2. A path-integral description of spin}
\medskip

To describe the dynamics on SU(2), let us work in a
specific matrix representation, the defining representation of
SU(2), and take for a basis
$T_i:=\dfrac{\s_i}{2i}$
for which $\tr(T_iT_j)=-\frac12\d_{ij}$ and
$[T_i\,,T_j]=\epsilon_{ijk}T_k$.
Then, for a free particle on
SU(2), the transition amplitude from an initial point $g_0$ at time
$t=0$ to a final point $g_1$ at time $t=T$ is represented by the
path-integral
$$
Z =
\int \D g\exp\biggl( -i\int_0^T dt\,\tr(g^{-1}\dot g)^2
\biggr)\,,
\eqn\pisu
$$
where $\D g :=\prod_tDg(t)$, and the formal sum is over
all possible paths $g(t)$
such that $g(0)=g_0$ and $g(T)=g_1$.
Conditions (\conditions) suggest that, in order to recover the
path-integral description of spin, we must impose
constraints associated with the right action, and for this
purpose we  consider the phase space version of
(\pisu),
$$
Z =
\int \D g\D\R\exp\biggl(i\int_0^T dt\,\(2\tr
\R (g^{-1}\dot g)+\tr\R^2\)\biggr)\,.\eqn\pspisu
$$
The measure for the momentum variables is taken to
be $\D\R=\prod_t\prod_{i=1}^3d\R_i(t)$
where the components  $\R_i:=-2\tr(\R T_i)$  are
the classical right momenta which generate the right action on SU(2) (more
details on this identification will be presented in Sect.\thinspace4 and~5,
{}for now
all we really need to note is that by integrating out the momentum
in (\pspisu) one recovers (\pisu)).

We now need to determine the  effective
classical counterpart to the conditions
(\conditions) to be used in the path-integral description of spin.
We cannot simply deduce that $\R_3=j$ and $\R_+=0$,
since $\R_+$ is not an element of su(2). However, taking expectation
values in the states (\spinstates) we can deduce that $\<\widehat\R_3\>=j$,
while $\<\widehat\R_1\>=\<\widehat\R_2\>=0$. This suggests that we
should take the effective classical version of (\conditions) to be
$$
\R_3=j\qquad\hbox{and}\qquad \R_1=\R_2=0\,.\eqn\cconditions
$$
The Poisson bracket relations among the right momenta are
$\{\R_i\,,\R_j\}=\epsilon_{ijk}\R_k$ (for this derivation, see
Appendix B), hence, from (\cconditions),
we see that (for $j\ne0$)
these represent a mixed set of constraints with $\R_3-j$
being first class while $\R_1$ and $\R_2$ are second class.
{}From this it is clear that the classical set of constraints
(\cconditions) leads
to the quantum conditions (\conditions) since these represent
the
\lq maximal first class subalgebra' constructed out of the
constraints (\cconditions) (modulo an ambiguity in choosing the set).
If $j=0$
then all the constraints are first class and we reduce to a point,
which is consistent with the quantum theory where the states are
just the constant functions on SU(2).

Restricting to the situation when $j\ne0$, we note that the first
class constraint function $\R_3-j$ will generate a U(1)-action
$$
g \longrightarrow g\,s\,,
\qquad{\rm where}\qquad s=e^{\theta T_3}\,,
\eqn\usk
$$
as a gauge symmetry.  Combining with
the second class constraints, we see that the original phase space
$T^*{\rm (SU(2)) = SU(2)} \times \Real^3$ is now reduced to
the {\it compact} phase space ${\rm SU(2)/U(1)} \simeq S^2$.
It is this compactness of the phase space that
results in the finite dimensional Hilbert space; indeed, the phase
space $S^2$ was postulated from the outset by Nielsen and Rohrlich
in their path-integral description of spin.

A path-integral account of the system (\pspisu), subject to the
constraints (\cconditions), is now quite straightforward to develop. The
constraints (\cconditions)
imply that the measure in (\pspisu) is modified [\FS] resulting in the
constrained path-integral $$ \eqalign{ Z_{\rm spin}=\int \D
g\D\R\,\d(\R_1)\d(\R_2)&\d(\R_3-j)\d(\xi)|\{\R_3,\xi\}|\cr
&\times\exp\biggl( i\int_0^T dt\,\(2\tr\R (g^{-1}\dot
g)+\tr\R^2\)\biggr)\,,}\eqn\no $$ where $\xi = 0$ is some gauge
{}fixing condition for the constraint $\R_3=j$ which, for simplicity,
is taken to be a function of the configuration space variables only
so that the factor $|\{\R_3,\xi\}|$ be independent of $\R_i$.
Performing the momentum integrals then yields $$ Z_{\rm spin}=\int
\D g\,\d(\xi)|\{\R_3,\xi\}| \exp\biggl (i \int_0^T dt\, 2j \tr
T_3(g^{-1}\dot g)\biggr)\,. \eqn\stwopi $$

Changes in the gauge fixing are implemented
by the U(1)-action (\usk).  When applying the gauge
transformation $g(t)\to g(t)s(t)$ within the path-integral we must
preserve the
boundary conditions on the paths.  This implies $s(T)=s(0)=1$ or,
in view of the basis we are using,
$$
\theta(T) - \theta(0) = 4\pi n\,,
\eqn\condtheta
$$
for some integer $n$.  On the other hand,
under the transformation (\usk) the Lagrangian in the path-integral
(\stwopi),
$$
L_{\rm spin} = 2j\tr T_3 (g^{-1}\dot g)\,,\eqn\stwolag
$$
changes to
$$
L_{\rm spin} \longrightarrow L_{\rm spin}-j\dot \theta\,.
\eqn\no
$$
Thus, in order for the path-integral (\stwopi) to be independent of the
gauge fixing condition $\xi$ (within the class of such gauge fixing
connected by the transformation (\usk))\note{
There is a subtlety with the gauge independence of the path-integral
due to the existence of a Gribov ambiguity; however, we
shall not address this point here (see [\David]).}
we must ensure that $e^{-i\int dt \, j\dot \theta} = 1$, or
$$
j (\theta(T)-\theta(0)) = 2\pi \times \hbox{integer}\,.
\eqn\jcondition
$$
{}For this to be true for any $n$ in (\condtheta)
we must have $2j \in \Z$, and hence we have
recovered the quantization condition for spin.  (Recall
that there is a two to one correspondence between SU(2) and SO(3),
in particular $1_{{\rm SO(3)}}=\{1,-1\}_{{\rm SU(2)}}$.  Hence, had
we been working on SO(3), the condition (\condtheta) would be
replaced by
$\theta(T) - \theta(0) = 2\pi n$,
and we would recover the integer quantization for spin $j \in \Z$
in that case.)

A useful local set of coordinates on SU(2) are the Euler angles
$(\a,\b,\gamma)$,
where $g=e^{\a T_3}e^{\b T_2}e^{\gamma T_3}$
($0\le\a<2\pi$, $0\le\b\le\pi$,
$0\le\gamma<4\pi$). In terms of these the path-integral (\stwopi) becomes
$$
Z_{\rm spin}=\int \D \a\D \b\D \gamma\,\sin\b\,
\d(\xi)\,\left|\frac{\partial\xi}{\partial\gamma}\,\right|\exp
\biggl( -i\int_0^Tdt\,j
( \dot\a \cos\b + \dot\gamma)\biggr)\,.
\eqn\localstwopi
$$
A gauge fixing that is only ill defined at the south pole ($\b=\pi$)
is given by
$$
\xi_+=\cases{\gamma+\a,&if $\,0\le\gamma<2\pi$;\cr
             \gamma-2\pi+\a,&if $\,2\pi\le\gamma<4\pi$.\cr
}
\eqn\xiplus
$$
This leads to the path-integral
$$
Z_{\rm spin}=\int\D \a\D \b\,\sin\b\exp\biggl(
i\int_0^Tdt\,j \dot \a
(1 - \cos\b)\biggr)\,.
\eqn\nrpi
$$
The Lagrangian $j \dot\a(1 - \cos\b)$ is well defined (actually
zero) at the north pole, while at the south pole it becomes
$2j\dot\a$.  Paths going $n$-times around this singular
point contribute $2j \times 2\pi n$ to the action, which is an irrelevant
phase since $2j$ is an integer. This result is also expected since
one can use the gauge invariance (\usk) to go from the gauge
{}fixing condition (\xiplus) to
$$
\xi_-=\cases{\gamma-\a,&if $\,0\le\gamma<2\pi$;\cr
             \gamma-2\pi-\a,&if $\,2\pi\le\gamma<4\pi$,\cr
}
\eqn\ximinus
$$
which is only ill defined at the north pole.

Expression (\nrpi) is equivalent to that proposed by Nielsen and
Rohrlich [\NR] (see also [\Fadetal])
{}for a path-integral description of spin. Here we have
seen that it can be derived from a constrained path-integral
{}formulation on SU(2).  The point to be observed is that
the above construction admits an interpretation in terms of the
generalized Dirac approach.  Indeed, the conditions (\conditions)
on which our constrained path-integral is based are of
the type (\npstates) and hence, following the generalized Dirac
approach, they represent the allowed anomalous behaviour for this
system starting with the (classical) first class
constraints given by (\cconditions) with $j = 0$, which are of the
type (\con).  Furthermore, the quantization of spin $j$ is
correctly deduced from the consistency of the path-integral.
In other words, using the generalized Dirac
approach one can construct the quantum mechanics of spin
by quantizing on the trivial coset ${\rm G/H = SU(2)/SU(2)}$.
It is now evident that,
reinstating the abelian group $\Real^3$ to turn SU(2) into
${\rm \widetilde E(3)}$, we can obtain spin
(in addition to the $\Real^3$ degrees of
{}freedom) by quantizing on $Q = \Real^3$ identified with the coset
$\widetilde{{\rm E(3)}}$/SU(2),
which is the result Mackey obtained in his approach using the
system of imprimitivity based on $\widetilde{{\rm E(3)}}$/SU(2).
In this case, the corresponding
path-integral describing a particle moving in $\Real^3$ with
spin has an effective phase space $T^*\Real^3\times S^2$, where the
two-sphere has radius $j^2$.

We have seen that the classical
spinless particle, described by the first class constraints
(\cconditions) with $j=0$, can acquire spin upon quantization by
the change in the structure of the constraints, which is {\it
required}
if we are to recover Mackey's result.
Such a change in the
constraints is usually anathema to the whole constrained formalism;
indeed the avoidence of this is one of the clearest ways of
stating the apparent need to cancel gauge anomalies in
Yang-Mills theory with chiral fermions.
However, we have seen through this specific example that such a
metamorphosis of constraints can indeed happen ---
with important physical consequence.
Before musing on
the possible implications of this for gauge theories,  we first need
to show that this flexibility in the class of the constraints is
generic; that is, we shall demonstrate that
such a generalized Dirac approach
is needed in order to recover Mackey's description of the
inequivalent quantizations on a generic G/H.
\vfil\eject

\secno=4 \meqno=1

\noindent
{\bf 4. Generalized Dirac approach to quantizing on coset spaces}
\medskip
\noindent
We have seen in the previous section that a constrained description
of Mackey's account of spin can be developed which has many
attractive features; not the least of which is the avoidence of
vector-valued wave functions in the  characterization of quantum
states. The
aim of this section is to extend that constrained analysis to the
general coset space G/H discussed by Mackey. We will clearly state
the generalized version of Dirac's approach appropriate
to this situation, and prove that the results so obtained agree with
those proposed by Mackey for the different quantum sectors of these
theories.  Finally we shall investigate the group invariance of the
physical states, and show that they are defined over a product of
the coset space G/H and a coadjoint orbit of H.

\bigskip
\noindent
{\it 4.1.
Hamiltonian dynamics on {\rm G} and classical reduction to} G/H
\medskip

The systems we wish to quantize are those describing free (geodesic)
motion on the configuration space $Q \simeq {\rm G/H}$,
with respect to the metric
$g_{\a\b}$ induced from the Killing metric on G.
This dynamics, in turn, can be recovered from a
reduction of the free motion on the extended configuration space G.
Here we want to recapitulate
how this reduction is described within the
Hamiltonian formulation of dynamics.

The kinematical arena for the Hamiltonian description of dynamics on
the Lie group
G is the phase space $T^*$G [\AM]. This cotangent bundle is actually a
trivial bundle over G and can be identified with ${\rm G} \times \g$.
If $\{T_m\}$
is a basis of $\g$
so that the point $\R\in\g$ can be written as
$\R=\R^mT_m$, and $g^{-1}dg$ is the left-invariant Maurer-Cartan 1-form
on G, then the above trivialization of $T^*$G amounts to identifying
$(g,\R)\in{\rm G}\times\g$ with $\R_m(g^{-1}dg)^m\in T^*$G, where
$\R_m:=\Tr(T_m\R)$.

As with any cotangent bundle, this phase space comes equipped with a
canonical symplectic 2-form $\omega$ from which the Poisson bracket
between functions can be calculated. In terms of the above
trivialization of $T^*$G, this symplectic 2-form is given by
$$
\omega= d\theta\,,
\qquad\hbox{where}\qquad
\theta := - \Tr \R (g^{-1}dg)\,.
\eqn\symplecticG
$$
Using a matrix representation of the elements of G so that
$g\in {\rm G}$
has matrix elements $g_{ij}$, this symplectic form generates the
{}fundamental Poisson bracket (see Appendix~B):
$$
\{g_{ij}\,, g_{kl}\} = 0\,,
\qquad
\{\R_m\,, g_{ij}\} =(gT_m)_{ij}\,,
\qquad
\{\R_m\,, \R_n\} =f^l_{mn}\R_l\,.
\eqn\pbs
$$
Thus we see that the \lq right-currents' $\R_m$ generate
the right action of G on itself,
$g \rightarrow g \tilde g$,
where $ \tilde g\in\,$G.
There is also a left action of G on itself,
$g \rightarrow \tilde g^{-1} g$,
which is generated by the \lq left-currents' $\L_m$ given by
$$
\L_m := -\Tr(T_mg\R g^{-1})\,.
\eqn\lr
$$
The left-currents satisfy the Poisson bracket
$$
\{\L_m\,,g_{ij}\} =-(T_mg)_{ij}\,,
\qquad
\{\L_m\,,\L_n\} =f^l_{mn}\L_l\,,
\eqn\pbsl
$$
which can be confirmed from (\pbs) and (\lr).
It is clear from the left and the right group actions that the
left and right-currents commute,
$$
\{\R_m\,,\L_n\}=0\,.
\eqn\pbsc
$$

We take as our Hamiltonian on G
$$
H_{\rm G} = \tfrac12\Tr \R^2 = \tfrac12\eta^{mn} \R_m \R_n\,,
\eqn\hamonG
$$
which can equally be written, from (\lr), as $H_{\rm G}
= \tfrac12\Tr \L^2$.
The equations of motion derived from (\hamonG) are
then
$$
\frac{d\ }{dt}\(g^{-1}\dot g\)=0\,,\eqn\eqmoG
$$
which decribe geodesic motion on G.  To confirm this, we
introduce a set of local coordinates $\{x^\mu\}$
($\mu = 1, \ldots, \dim\,$G)
parametrizing $g = g(x)$, and recall
that the natural metric on G is given by
$g_{\mu\nu} := \eta_{mn}\, v_{\,\,\mu}^m\, v_{\,\,\nu}^n$,
where $v = v^m_{\,\,\mu} \, dx^\mu \, T_m := g^{-1}dg$
is the vielbein 1-form defined from the left invariant 1-form.
Then, using these, we can convert (\eqmoG) into
$$
\ddot x^\mu + \Gamma^\mu_{\nu\lambda}(x)\, \dot x^\nu
\dot x^\lambda = 0\,,
\eqn\geoG
$$
where $\Gamma^\mu_{\nu\lambda}$
is the Levi-Civita connection.
An alternative and quicker way to arrive at the geodesic equation
(\geoG) is to derive the
Lagrangian from (\symplecticG) and (\hamonG) by Legendre transform,
$$
L_{\rm G} = \tfrac12 \Tr(g^{-1}\dot g)^2 = \tfrac12
g_{\mu\nu}(x)\, \dot x^\mu \dot
x^\nu\,,
\eqn\lagonG
$$
where now (\geoG) is obvious.

{}For the reduction G $\rightarrow$ G/H,
we first note that
the right action of the subgroup H on G is generated by the
right currents $\R_i=\Tr(T_i\R)$, where $\{T_i\}$
is a basis of $\h$ in the reductive
decomposition $\g=\h\oplus\r$.  The
currents $\R_i$ form the algebra $\h$ under Poisson bracket,
$$
\{\R_i\,,\R_j\}=f^k_{ij}\R_k\,,
\eqn\no
$$
and can be used to implement the reduction from the
extended phase space $T^*$G to the reduced phase space
$T^*$(G/H), {\it i.e,}
we take as our classical, first class constraints
$$
\R_i=0\,.
\eqn\constonG
$$
The Hamiltonian (\hamonG) preserves these constraints, and
on the constrainted surface given by (\constonG) it
reduces to the (physical) Hamiltonian,
$$
H_{\rm G/H} = H_{\rm G}|_{\R_i=0}=\tfrac12\eta^{ab}\R_a\R_b\,.
\eqn\hamonGoH
$$
On the other hand,
given a section $\s:\,{\rm G/H \mapsto G}$
we can write
$g = \s \, h$ with
$\s \in {\rm G}$, $h \in {\rm H}$.
Using a set of
local coordinates $\{q^\alpha\}$ on G/H to parametrize $\s =
\s(q)$, and performing Legendre transform we obtain
the reduced Lagrangian
$$
L_{\rm G/H} = \tfrac12 \Tr (\s^{-1}\dot \s \vert_\r)^2
= \tfrac12 g_{\a\b}(q) \,\dot q^\a \dot q^\b \,,
\eqn\lagoncoset
$$
where we have defined
the vielbein $e = e^a_{\,\,\alpha} \, dq^\alpha\,
T_a := \s^{-1}d\s\vert_\r$ to provide the metric
$g_{\a\b} := \eta_{ab}\, e_{\,\,\a}^a\, e_{\,\,\b}^b$
on G/H.  Then again, it is obvious that the equations of motion
derived in the reduced system are of the form,
$$
\ddot q^\a+\Gamma^\a_{\b\gamma}(q)\, \dot q^\b\dot q^\gamma = 0\,,
\eqn\no
$$
where $\Gamma^\a_{\b\gamma}$ is the Levi-Civita connection
constructed out of the metric $g_{\a\b}$.
Thus we see that the dynamics of the reduced
system is indeed that describing geodesic motion on G/H.

\bigskip\noindent
{\it 4.2. Dirac's (naive) quantization on {\rm G/H}}
\medskip
The first step in any constrained description of quantizing on G/H
is to initially ignore the constraints (\constonG) and quantize on
the extended configuration space G. Any group G can be viewed
trivially as a coset space G/$\{1\}$, hence, using Mackey's approach
discussed in Sect.\thinspace2, there
is a well defined
quantization in which the states are simply the square integrable
wave functions $\psi(g)$ on G and the integration is with respect to the
Haar-measure on G. Then the representation (\induced) of G on these
states is simply the left-regular representation
$$
(U_{\rm L}(\tilde g)\psi)(g)=\psi(\tilde g^{-1}g)\,.
\eqn\lreg
$$
The generators of the left action are, from (\inducealgebra),
simply the quantized left-currents
$\widehat \L_m$, and can be identified with ($i$ times) the
right-invariant vector fields on G.
On top of this representation of G, there is also the right-regular
representation on these states
$$
(U_{\rm R}(\tilde g)\psi)(g)=\psi(g\tilde g)\,,\eqn\rreg
$$
which is clearly generated by the quantized right-currents
$\widehat\R_m$ and can be identified with ($-i$ times) the left
invariant vector fields on G.  Hence the commutators among the currents
are
$$
[\widehat \R_m\,, \widehat \R_n]= i f^l_{mn}\widehat \R_l\,,
\qquad
[\widehat \L_m\,, \widehat \L_n] = i f^l_{mn} \widehat \L_l\,,
\qquad
[\widehat \R_m\,, \widehat \L_n]= 0\,,
\eqn\qpbs
$$
which are consistent with the Poisson bracket relations
(\pbs), (\pbsl) and (\pbsc).

Dirac's approach to the constrained quantization on G/H
identifies the physical (reduced) wave functions with those wave
{}functions $\psi_\phy$ on G which satisfy
$$
\widehat\R_i\,\psi_\phy(g)=0\,.\eqn\diraccon
$$
Such a condition on states is taken to be the natural translation of
the classical condition\note{It should be noted, though, that in
the classical
theory gauge fixing is needed to directly isolate the physical states.
It is also the case that in the path-integral account of this
system one usually uses some gauge fixing.
So it is a peculiar artifact of this system
(based on the fact that
G is compact so that there are normalisable solutions
to (\diraccon) on $L^2(G)$) that gauge fixing appears to be
inessential in the operator formalism.
However, a more systematic approach [\David]
shows that gauge fixing has some significance even in this case.}
(\constonG).
Thus, for such physical states we have
$$
(U_{\rm R}(h)\psi_\phy)(g)=\psi_\phy(g)\,,\eqn\hinvar
$$
{}for all $h\in{\rm H}$ (or more precisely, for all $h\in{\rm H}$
connected to the identity $h = 1$).
Then from (\rreg) we find $\psi_\phy(g)=\psi_\phy(gH)$,
which implies that
the physical wave functions on G are, in effect, just
wave functions on G/H.
Since the left and right regular
representations commute, we see that the left-regular represention
 (\lreg) becomes the trivial
representation on $L^2({\rm G/H})$ in Mackey's account
of quantizing on G/H (see (\induced)).

On the extended state space the quantized Hamiltonian (\hamonG) is
given by
$$
\widehat H_{\rm G} =\tfrac12\Tr \widehat\R^2
=\tfrac12\eta^{mn}\widehat\R_m\widehat\R_n\,,
\eqn\hamhat
$$
which on the physical states (\diraccon) reduces to the quantized
version of the Hamiltonian (\hamonGoH),
$$
\widehat
H_{\rm G/H} =\tfrac12\eta^{ab}\widehat\R_a\widehat\R_b\,.\eqn\hamphyhat
$$
This is precisely the Laplace-Beltrami operator $-\frac12
\triangle_{\rm LB}$ acting on the wave function on G/H.

\bigskip
\noindent
{\it 4.3. Generalized Dirac quantization on {\rm G/H}}
\medskip

We have seen that  Dirac's (naive) approach to quantization
applied to the coset space G/H recoveres only the trivial sector in
Mackey's account of the inequivalent quantizations possible on such
a space. Thus, following the philosophy outlined in the
Introduction (as summarized in Fig.\thinspace2 there) we need to
generalize Dirac's reduction prescription (\diraccon) so that all
quantizations are recovered. From our analysis of Sect.\thinspace3,
we expect this to be achieved by putting the constraints
$\widehat\R_i$ equal to some constants, rather than zero as the
classical theory might suggest. Following our discussion of Mackey,
we expect these constants to be related to the irreducible
representaions of H. To start this generalization of Dirac's
approach, though, we first need the following technical
result that allows
us to disentangle the H and G/H parts of the wave functions on
G:
$$
L^2({\rm G})\simeq L^2({\rm G/H}
\times {\rm H})\simeq L^2({\rm G/H})\otimes
L^2({\rm H}).\eqn\wavefunctfact
$$
This is, in fact,
almost obvious if we recall that any fibre bundle is locally
trivial, and can be written as a product of the base space and the
{}fibre.  For a formal proof, let us choose a section $\s$ and
define the unitary mappings
$T:L^2({\rm G})\mapsto L^2({\rm G/H}\times{\rm H})$ and
$T^*:L^2({\rm G/H}\times{\rm H})\mapsto L^2({\rm G})$
by
$$
(T\psi)(q,h) = \psi\bigl(\s(q)h\bigr)\,,
\qquad
(T^*\psi)(g) = \psi\bigl({\rm pr}(g),
\s^{-1}({\rm pr}(g)) g\bigr)\,,
\eqn\no
$$
where pr$(g)$ is the projection ${\rm G \mapsto G/H}$ (see
Sect.\thinspace2.1).
Then it is easy to see that these mappings are the
inverse to each other; for example, we have
$$\eqalign{
(TT^*\psi)(q,h)&=(T^*\psi)\bigl(\s(q)h\bigr)\cr
&=\psi\bigl({\rm pr}(\s(q)h),
\s^{-1}({\rm pr}(\s(q)h))\s(q)h\bigr)\cr
&=\psi(q,h)\,,
}\eqn\no
$$
and the other way around is even easier.
The unitarity of these mappings follows from the definition (\measure)
of the invariant measure on G/H.  On account of this isomorphism,
in what follows we sometimes write $\psi(g)$ as $\psi(q,h)$ or vise versa
by abuse of notation.

On $L^2({\rm G/H}\times{\rm H})$ the left-regular representation
(\lreg) becomes
$$
(U_{\rm L}(g)\psi)(q,h)
=\psi\bigl(g^{-1}q,h_\s(g^{-1},q)h\bigr)\,,
\eqn\lregg
$$
where $h_\s(g^{-1},q)$ is the element of H defined in (\hsigma).
Given this decomposition of $L^2({\rm G})$ we can further decompose the
wave functions on H by using the Peter-Weyl theorem.
Recall that given an irreducible representation  of H,
labelled by the
heighest weight $\chi$, we introduce a basis of states
$\ket{\chi,\mu}$
and write the matrix elements of this representation as
$\pi^\chi_{\mu\nu}(h)$ as in (\matel).
Then we have the orthogonality relations
$$
\int_{{\rm H}}\!Dh\,
\pi^{\chi*}_{\mu\mu'}(h)\,\pi^\chi_{\nu\nu'}(h)=
\frac1{d_\chi}\d_{\mu\nu}\d_{\mu'\nu'}\,,
\eqn\no
$$
where $d_\chi$ is the dimension of the representation
$\chi$, and the Peter-Weyl theorem
tells us that we can write any wave function $\psi(h)$ on H as
$$
\psi(h)=
\sum_{\chi,\mu,\mu'}C^\chi_{\mu\mu'}\pi^{\chi*}_{\mu\mu'}(h)\,.
\eqn\no
$$
Combining this result with the isomorphism (\wavefunctfact), we see
that any wave function $\psi(g)$ on G can be written as
$$
\psi(g)=\sum_{\chi,\mu,\mu'}\psi^\chi_{\mu\mu'}(q)
\pi^{\chi*}_{\mu\mu'}(h)\,.\eqn\no
$$
where now the coefficients $\psi^\chi_{\mu\mu'}(q)$ are wave
{}functions on G/H.

Now let $K$ be an element of the Lie algebra $\h$ and consider the
operators
$$
\widehat\phi_i = \Tr T_i(\widehat\R-K) = \widehat\R_i-K_i\,.
\eqn\newcons
$$
We identify
these with the new, effective constraints of the quantum
theory. However, the functions $\widehat\phi_i$ represent a
mixed set (first and second class) and
thus cannot be directly used to define the physical states.
Indeed, from (\qpbs) the commutator of these operators are
$$
[\widehat\phi_i\,, \widehat\phi_j] =
if^k_{ij}\,\widehat\phi_k+i\Tr([T_i,T_j]K)\,,
\eqn\no
$$
and hence
the $\Tr([T_i,T_j]K)$ term signals the existence of some second
class components in the set (\newcons).
In order to isolate the first class subset of (\newcons) we
consider the subalgebra ${\frak s}_K$
given by the kernel of the adjoint action of $K$ in $\h$,
$$
{\frak s}_K := {\rm Ker(ad}_K) \cap \h\,.
\eqn\liesk
$$
{}For a generic $K$, that is, if $K$ is a {\it regular
semisimple} element in $\h$,
the subalgebra ${\frak s}_K$ is precisely the Cartan
subalgebra ${\frak t}$ of $\h$
containing $K$ [\Humphreys].
If not, ${\frak s}_K$ is
larger than ${\frak t}$ and, due to the non-degeneracy of ${\frak
t}$ with respect to the Killing form, admits the
decomposition ${\frak s}_K = {\frak t} \oplus {\frak c}$ where
${\frak c}$ is the orthogonal complement.
Choosing a basis $\{T_s\}$ in ${\frak s}_K$ (we shall use $\{T_{r}\}$
and $\{T_{p}\}$ for the bases in ${\frak t}$ and ${\frak c}$,
respectively),
we see that for
any $T_j \in \h$ we have $\Tr ([T_s, T_j] K)=0$ and hence
the first class components in (\newcons) are given by
$$
\widehat\phi_s := \Tr T_s (\widehat \R - K)\,.
\eqn\no
$$
Conversely, from the semisimplicity of $\h$ it follows
that these $\widehat \phi_s$ form
the maximal set of the first class components in
(\newcons).

Just as we did for the spin example in Sect.\thinspace3, we now need
to find a \lq maximal first class subalgebra' formed by the
operators in (\newcons),
which involves working with the complex extension $\h_{\rm c}$
of the algebra $\h$.
Introducing a Chevalley basis $(H_{\a_r}, E_{\pm\varphi})$
in $\h_{\rm c}$ (see Appendix~A) where we take $T_r :=
\frac{1}{i}H_{\a_r}$ ($r = 1, \ldots,
{\rm rank\,H}$),
we see that a maximal first class subalgebra of (\newcons) is given by
the set $(\widehat\phi_s ,\widehat\phi_{\varphi})$, where we
have defined
$$
\widehat\phi_{\varphi} = \Tr E_{-\varphi}(\widehat \R-K)\,,
\eqn\no
$$
for all positive roots $\varphi$.
If $K_r = \Tr(T_r K)$
has integer values corresponding to the heighest weight
$\chi$, {\it i.e.}, $K_r = \chi(H_{\alpha_r})$,
then the physical states defined by the conditions
$$
\widehat\phi_s\, \psi_\phy(g) = 0 \qquad\hbox{and}\qquad
\widehat\phi_{\varphi}\, \psi_\phy(g) = 0\,,
\eqn\pscond
$$
have the solutions of the form
$$
\psi_\phy(g)
= \psi_\phy(q,h)= \sum_\mu\psi_\mu(q) \pi^{\chi*}_{\mu\chi}(h)\,.
\eqn\physk
$$
{}For the generic case where $K$ is regular semisimple
the coefficient functions $\psi_\mu(q)$
are completely arbitrary, while for non-generic $K$ they must satisfy
certain conditions so that the physical wave functions (\physk)
be invariant under the transformations generated by $\widehat
\phi_p$.
In any case the wave functions (\physk) provide
the correct identification with the physical
states corresponding to the quantization on G/H labelled by $\chi$
in the sense that the coefficient wave functions in (\physk) are
precisely the vector-valued wave
{}functions (\vvwf) used in Mackey's analysis.
Indeed, given a
physical wave function (\physk), the coefficient functions in
(\physk) are obtained as
$$
\psi_\mu(q)
= d_\chi\int_{\rm H}\!Dh\,
\psi_\phy(q,h)\pi^\chi_{\mu\chi}(h)\,.
\eqn\coeff
$$
Hence under the left action (\lregg) we have
$$\eqalign{
\psi_\mu(q)
& \to
d_\chi\int_{\rm H}\!Dh\,
\psi_\phy \bigl(g^{-1}q,h_\s(g^{-1},q)h\bigr)
\pi^\chi_{\mu\chi}(h)\cr
&= d_\chi\int_{\rm H}\!Dh\,
\sum_\nu\psi_\nu(g^{-1}q)\pi^{\chi*}_{\nu\chi}
(h_\s(g^{-1},q)h)
\pi^\chi_{\mu\chi}(h)\cr
&= d_\chi \sum_{\mu,\nu}
\psi_\nu(g^{-1}q) \pi^{\chi*}_{\nu\lambda}
(h_\s(g^{-1},q))\int_{\rm H}\!Dh\,
\pi^{\chi*}_{\lambda\chi}(h)\pi^\chi_{\mu\chi}(h)\cr
&= \sum_{\nu}\pi^\chi_{\mu\nu}(h_\s(g,g^{-1}q))\psi_\nu(g^{-1}q)\,.
}\eqn\no
$$
Thus we see that the left-regular representation (\lregg)
reproduces exactly the induced representation (\induced), and hence
$\psi_\mu(q)$ are identified with the vector-valued wave functions
used by Mackey.  This proves the equivalence between
our generalized Dirac approach and
Mackey's approach,
and the formulae, (\physk) and (\coeff), provide the mapping
between the two.

In the trivial sector we have seen from (\hinvar)
that the physical wave functions on G can be identified with the
wave functions on the physical (classical) configuration space G/H.
In the remainder of this section
we shall investigate which space the wave functions (\physk) are
naturally defined over when we are not in the trivial sector.

The first thing to note is that the physical states
(\physk) are {\it not}
H-invariant. In fact, using the identification (\wavefunctfact),
the right action of H on $L^2({\rm G})$ is simply
$$
(U_{\rm R}(\tilde h)\psi)(q,h)=\psi(q,h\tilde h)\,.\eqn\haction
$$
So acting on a physical state (\physk) we get
$$\eqalign{
(U_{\rm R}(\tilde h)\psi_\phy)(q,h)&=\psi_\phy(q,h\tilde h)\cr
&=\sum_\mu\psi_\mu(q)\pi^{\chi*}_{\mu\nu}(h)\pi^{\chi*}_{\nu\chi}
(\tilde h)\,,
}\eqn\hphyss
$$
which, for a
general $\tilde h\in\,$H, is not related in any obvious way to
the original physical state. However, if $\tilde h = s \in
\SK$
where $\SK$ is the subgroup of H obtained by the exponential mapping
of the subalgebra ${\frak s}_K$ in (\liesk),
then we can write $s=e^{T_r\theta^r} e^{T_{p}\theta^{p}}$,
whereby (\hphyss) becomes
$$
\psi_\phy(q, hs)=e^{iK_r \theta^r}\psi_\phy(q,h)\,.
\eqn\no
$$
{}From (\haction) we see that this $\SK$-invariance of the physical states
only affects the H~factor in the decomposition
$L^2({\rm G}) \simeq L^2({\rm G/H}\times{\rm H})$.
Indeed, just as we did for wave functions on G, we can
decompose further $L^2({\rm H})$ as
$$
L^2({\rm H}) \simeq L^2(O_K \times \SK)\,,
\eqn\no
$$
where $O_K:={\rm H}/\SK$ is the coadjoint orbit through $K\in\h$.
Clearly, due to the
$\SK$-invariance the physical wave functions on G can be
identified with wave functions on the (effective) physical space
${\rm G/H} \times O_K$.  More precisely, these wave functions
are associated
with quantizing on the classical phase space $T^*{\rm (G/H)} \times
O_K$ --- they are {\it not}
the whole of $L^2({\rm G/H}\times O_K)$.  Thus
we see that, in addition to the degrees of freedom on G/H that we
naively expect, we find extra degrees of freedom represented by
the coadjoint orbit $O_K$.  In the example we discussed in
Sect.\thinspace3,
these were in fact spin degrees of
{}freedom , and by this reasoning
they will be henthforth
called \lq generalized spin' in the general case.

The loss of the H-invariance, though,
on the extended state space is not too
attractive when we try and generalize these ideas to field theory
where we still want to have gauge invariance.
We may, however, restore the H-invariance
at the expense of working on a larger state space.
This can easily be accomplished by constructing new wave functions
defined on ${\rm G \times H}$
{}from the physical wave functions in (\physk) by
$$
\Psi (g, h) := \psi_\phy(gh)
= \sum_\mu \Psi_\mu (g) \pi^{\chi*}_{\mu\chi}(h)\,,
\eqn\physonG
$$
where the coefficient wave functions on G are given by
$$
\Psi_\mu(g) := \sum_\nu \psi_\nu ({\rm pr}(g))
\pi^{\chi*}_{\nu\mu} (\s^{-1}({\rm pr}(g)) g)\,.
\eqn\waveonG
$$
Clearly, $\Psi(g,h)$ has the (trivial) H-invariance as well as the
$\SK$-invariance,
$$
\Psi(g\tilde h, \tilde h^{-1} h) = \Psi(g,h)\,,
\qquad
\Psi(g, h s) = e^{i K_r \theta^r} \Psi(g,h)\,.
\eqn\no
$$
Note that the coefficient wave functions (\waveonG)
satisfy the identity
$$
\Psi_\mu(g\tilde h) = \sum_\nu \pi^\chi_{\mu\nu}
(\tilde h^{-1}) \Psi_\nu(g)\,.
\eqn\macid
$$
Conversely, given a collection
of wave functions $\Psi_\mu(g)$ on
G which satisfies (\macid), then they can be written in the form
(\waveonG).  In fact these are the form of the vector-valued wave
{}functions used by Mackey [\Mackey].
\vfil\eject

\secno=5 \meqno=1

\noindent
{\bf 5. Path-integral description for quantizing on {\rm G/H}}
\medskip
\noindent
In the previous section we have presented a generalization of
Dirac's approach to constrained quantization, in order to deal with
inequivalent quantizations on a coset space G/H.
In this section we
develop the corresponding language in the path-integral
which renders the treatise more transparent and
admits a more intuitive understanding of the
inequivalent quantizations.
In particular,
we shall see that in the path-integral
the appearance of the generalized
spin discussed in Sect.\thinspace4 is realized immediately,
and that the emergence of the H-connection is also trivial to
observe.
{}Furthermore,
the consistency of the path-integral under
the $\SK$-transformations requires
the parameters in $K$ to be precisely those labelling
the irreducible representations $\chi$ of H.
The equations of motion on G/H turn out to be
of the type of the
Wong equations [\Wong], describing a particle minimally coupled to the
H-connection and the generalized spin.

\bigskip
\noindent
{\it 5.1. Lagrangian realization in the path-integral}
\medskip

We have seen, in Sect.\thinspace4, that the inequivalent
quantizations on the coset space G/H can be derived from a
generalized form of Dirac's constrained quantization applied to the
classical system defined on G, subject to the (classical) first
class constraints $\R_i=0$. This account has been an operator one. To
develop the corresponding path-integral version we will simply
extract from that analysis the fact that in the quantum theory these
classical constraints become the effective, anomalous constraints
$$
\phi_i = \Tr T_i (\R - K) = \R_i - K_i = 0\,,
\eqn\const
$$
where $T_i \in \h$ (for our notational conventions, see Appendix A),
and then use the familiar prescriptions for implementing these
within a path-integral [\FS].
At this stage the $K_i$ are only assumed to be arbitrary constants.

Due to the existence of a first class subset of these constraints
(\const) given by
$$
\phi_s = \Tr T_s (\R - K) = 0\,,
\eqn\fconst
$$
{}for $T_s \in {\frak s}_K = {\rm Ker(ad}_K) \cap \h$,
the resulting theory acquires a gauge symmetry
under the $\SK$-transformations associated with the algebra ${\frak
s}_K$.
Accordingly, we introduce gauge fixing conditions $\xi_s = 0$
so that the total set of
constraints $\varphi_k := (\phi_i, \xi_s)$ is second class.
With this second class set of constraints the
phase space path-integral reads
$$
Z =
\int \D g \, \D \R \,\d(\varphi_k)\,
{\rm det}^{\frac12}\vert \{ \varphi_k, \varphi_{k'} \} \vert\,
\exp \biggl(i\int \theta - i \int_0^T dt\, H_{\rm G} \biggr)\,,
\eqn\ppi
$$
where $\theta$ is the canonical 1-form in (\symplecticG),
$$
\int \theta = - \int_0^T dt \, \R_m (g^{-1} \dot g)^m \,,
\eqn\no
$$
and $H_{\rm G}$ is the Hamiltonian (\hamonG).
The path-integral measure in (\ppi) is formally defined from
the volume (Liouville) form of the phase space $\omega^N$
($N = {\rm dim \, G}$) by taking its product over $t$,
$$
\D g \,\D \R = \prod_t \omega^N(t)\,, \qquad{\rm where}\qquad
\omega^N = \prod_{m = 1}^N  (g^{-1} dg)^m\, d\R_m\,.
\eqn\measure
$$
On account of the fact that the
operators in (\pscond) form the maximal first class subalgebra of
(\newcons), the operator version of (\const),
it is clear that the constrained path-integral (\ppi)
leads to the quantum theory equivalent to
that obtained by the operator description of Sect.\thinspace4.

Since the constraints (\const)
are (at most)
linear in the  momentum variables $\R_i$, we may trivially
implement them by integrating over all the momentum variables
$\R_m$.
{}For this, we first note that
the determinant factor in (\ppi) is proportional to
${\rm det}\vert \{ \phi_s\,, \xi_{s'} \} \vert$
on the constrained surface.  Thus, if we choose
the gauge fixing conditions $\xi_s = 0$ so that the determinant
be independent of $\R_m$, we can at once carry out the integrations on
$\R_m$ to obtain
$$
Z = \int \D g \, \d(\xi_s)\,
{\rm det} \vert \{ \phi_s, \xi_{s'} \}
\vert\, \exp\biggl(i\int_0^T dt\, L_{\rm tot}\biggr)\,,
\eqn\cpi
$$
where
$$
L_{\rm tot} = \frac 1 2 \Tr (g^{-1} \dot g \vert_{\r})^2
-  \Tr K (g^{-1} \dot g\vert_{\h})\,.
\eqn\lag
$$
Note that since the first term in the Lagrangian (\lag) is invariant
under $g \rightarrow g\,\tilde h$ for $\tilde h \in {\rm H}$
(which follows from the identity
$
h^{-1} X \,  h \vert_\r =
h^{-1} X\vert_\r \,  h
$
valid for any $ X \in \g$ as a consequence of the reductive decomposition),
it is actually   identical to
the Lagrangian (\lagoncoset) for
G/H.
On the other hand, it will be shown that
the second term in the Lagrangian (\lag)
represents the effects of non-trivial quantizations,
containing the H-connection and the generalized spin.
In passing we point out that the path-integral (\cpi)
is not quite a
configuration path-integral due to the second term which is first
order in time derivative.
We also
note that the measure $\D g =  \prod_{t,m} (g^{-1} dg)^m(t)$ in
(\cpi) is
{}formally the product of the Haar measure of the group G over $t$,
and by this expression it is meant to give
the sum over all possible paths allowed by the prescribed boundary
condition (to be discussed later), which includes \lq multi-winding'
paths going
around several times the fundamental region of the group manifold.

\bigskip
\noindent
{\it 5.2. Parameter quantization of $K$}
\medskip

We know, from Sect.\thinspace4, that the parameters in $K$ are
not quite arbitrary; they must be those (integers) that label the
highest weight
representation $\chi$ of H on which the Hilbert space is constructed.
We now show that this particular information
has already been incorporated in our path-integral.
To this end, we first observe that, under
the $\SK$-transformations
$g \rightarrow g \, s$ for $s \in \SK$
generated by the first class set
(\fconst),
the total Lagrangian varies as
$$
L_{\rm tot} \longrightarrow L_{\rm tot} + \Delta L_{\rm tot}\,,
\qquad \hbox{where} \qquad
\Delta L_{\rm tot} = -  \Tr K (s^{-1} \dot s)\,.
\eqn\ski
$$
As in the previous section, if we use
the parametrization
$s = e^{\theta^r T_r} e^{\xi^p T_p}$ where
$\{ T_r \}$ and $\{ T_p \}$ are bases in ${\frak t}$ and
${\frak c}$ in the orthogonal decomposition
${\frak s}_K = {\frak t} \oplus {\frak c}$,
then
$$
\Delta L_{\rm tot} = - {d\over{dt}}(K_r \theta^r)\,.
\eqn\td
$$
Thus the Lagrangian is invariant up to a total time derivative,
implying that
the equations of
motion are invariant under the $\SK$-transformation, which is
the $\SK$-symmetry at the classical level.

However, if we require the $\SK$-symmetry to persist
at the quantum level (which we must to ensure that the path-integral
(\cpi) is independent of the gauge fixing), then we need to
take into account the
boundary effect in the path-integral.  To examine this explicitly,
consider the transition amplitude from an initial
point $g_0$ at $t = 0$ to a final point $g_1$ at $t = T$.
The sum in the path-integral contains
all possible paths $g(t)$ going from $g(0) = g_0$ to
$g(T) = g_1$, but
to each such path there is a class
of paths related each other by a
gauge transformation,
$$
g(t) \longrightarrow g(t)\, s(t)\,, \qquad {\rm with} \qquad
s(0) = s(T) = 1\,.
\eqn\bdc
$$
The gauge invariance at the quantum level
requires that the paths within a
gauge equivalent class must contribute
to the sum of the path-integral with the same amplitude, {\it i.e.},
they must have the same phase factor.
Using the Chevalley basis for
our basis
$T_r = \frac1i H_{\alpha_r}$, we find from the periodic property
(see Appendix A) that gauge
transformations satisfying the condition (\bdc) are of the type,
$$
s(t) = e^{\theta^r(t) T_r}\,, \qquad{\rm with}\qquad
\theta^r (T) - \theta^r(0) = 2\pi n_r \,,
\eqn\parint
$$
where $n_r$ are integers\note{More precisely, $n_r$ are integers for
the universal covering group $\widetilde {\rm H}$ of H.
{}For a non-simply connected group H, they are multiple of integers
determined by the discrete normal subgroup
N of $\widetilde {\rm H}$ for which
H $\simeq \widetilde{\rm H}/{\rm N}$.
{}For instance, for SU(2) they are integers but for SO(3) $\simeq$
SU(2)/$\Z_2$ they are even integers.  Thus for non-simply connected
groups the quantization
condition for $K_r$ should be modified accordingly.}.
{}From this one sees that the requirement
$e^{i\int_0^T dt\, \Delta L_{\rm tot}} = 1$, or
$$
\int_0^T dt \, \Delta L_{\rm tot} = - 2\pi n_r\, K_r = 2\pi \times
\hbox{integer}\,,
\eqn\quant
$$
{}for any class of gauge transformations ({\it i.e.}, for any
$n_r$) is equivalent to
$$
K_r \in \Z\,, \qquad {\rm for} \quad r = 1, \ldots,
\hbox{rank}\,{\rm H}\,.
\eqn\pq
$$
These integer parameters
correspond precisely to the ones which label
the highest weight representations of H, with
the identification being $K_r = \chi(H_{\alpha_r})$, and
this parameter quantization of $K$
completes the bridge between the
operator description of Sect.\thinspace4 and the path-integral
description given
here.  Accordingly,
we see that our simple implementation of
the generalized Dirac approach in the path-integral,
where one just replaces the naive constraints (\constonG) with
(\const), results in the same quantizations as
those obtained in
Mackey's approach which is based on the system of imprimitivity on
G/H.

\bigskip
\noindent
{\it 5.3. Emergence of the generalized spin and the {\rm
H}-connection}
\medskip

We now examine the dynamical implications of the Lagrangian (\lag).
In particular,
we are interested in the effects which arise
when non-trivial quantizations are adopted for quantizing
on the coset space G/H.  From the identification that the constant
$K$ labels the representation $\chi$, we expect that
those effects
take place with the \lq strength' determined by $K$.

Let us start by decomposing $g$ as
$$
g = \s \, h\,, \qquad {\rm with} \qquad
\s \in {\rm G}\,, \quad h \in {\rm H}\,,
\eqn\dec
$$
where $\s = \s(q)$ is a section ${\rm G/H \mapsto G}$, and
$\{q^\alpha\}$ is a set of local coordinates on G/H.
Then the Lagrangian (\lag) becomes the sum of three terms:
$$
\eqalign{
L_{\rm tot} &= L_{\rm G/H} + L_{O_K} + L_{\rm int} \cr
     &= \frac 1 2 g_{\alpha\beta}(q)\, \dot q^\alpha \dot q^\beta
- \Tr K (h^{-1} \dot h)
- \Tr \bigl(h K h^{-1} A_\alpha(q)\bigr)\, \dot q^\alpha \,.
}
\eqn\llag
$$
Besides the first term in (\llag), which is the Lagrangian $L_{\rm G/H}$
(\lagoncoset)
{}for a free particle moving on G/H, we now recognize two terms
which represent the effects of non-trivial
quantizations
and are both proportional to $K$.
The effects are two-fold: first,
there appears extra degrees of freedom
$h \in {\rm H}$ which, as we shall see in the next section,
correspond to the generalized spin.  Their dynamics is governed by
$L_{O_K}$, which is the first order Lagrangian
for the coadjoint orbit
$O_K = {\rm H}/\SK$ of the group H passing through $K$.
(Note that the Lagrangian $L_{O_K}$ is invariant under
the $\SK$-transformations
and hence the phase space for the generalized spin is given by H/$\SK$.)
The second effect appears in the last term
$L_{{\rm int}}$, where one finds the H-connection
$$
A := \s^{-1}d\s \vert_\h = A_\alpha dq^\alpha\,,
\eqn\hconnection
$$
coupled minimally to the particle and the generalized spin.
Note that if H is abelian then the coadjoint orbits $O_K$ are just
points and therefore there appears no generalized spin.  If, on the other
hand, H = G then there emerges no H-connection; only the generalized spin
could exist.  The spin example discussed
in Sect.\thinspace3 falls into the latter case (where ${\rm G/H =
SU(2)/SU(2)}$), while the example of the quantization on $S^2$ to be
discussed in the next section is in the former case (where ${\rm G/H =
SO(3)/SO(2)}$).

The effects of the inequivalent quantizations in the dynamics
can be seen in the
equations of motion.  In terms of the ($\SK$-gauge invariant) variables
$$
S := - h K h^{-1}\,,
\eqn\gspin
$$
the equations of motion for $h$ derived from the Lagrangian
(\llag) are the covariant constancy equations,
$$
D_t S := {{dS}\over{dt}} + [A_\alpha(q), S] \, \dot q^\alpha = 0\,.
\eqn\emh
$$
Whereas, for the particle motion we get
$$
\ddot q^\alpha +
\Gamma^\alpha_{\beta\gamma}(q)\, \dot q^\beta \dot q^\gamma
- g^{\alpha\beta}(q) \,S_i \, F^i_{\beta\gamma}(q)
\, \dot q^\gamma = 0\,,
\eqn\emq
$$
where $F := dA + A \w A$ is the curvature 2-form.
Eqs.~(\emh) and (\emq)
are essentially the Wong equations [\Wong]
under the special, background non-abelian potential, the
H-connection, with the couplings (the parameters in $K$)
in $S$ taking only discrete values (they are quantized).

\bigskip
\noindent
{\it 5.4. {\rm H}-gauge structure}
\medskip
The Wong equations were originally designed
to provide a gauge invariant minimal coupling
with a non-abelian potential, as a generalization of the minimal
coupling with an electromagnetic potential.  The same is true for our
Lagrangian (\llag); it is invariant under
the H-gauge transformation,
$$
A \longrightarrow {\tilde h}^{-1}A\,\tilde h + {\tilde h}^{-1} d\tilde h \,,
\eqn\hcon
$$
which is induced by the change of section,
$$
\s \longrightarrow \s \, \tilde h\,, \qquad
h \longrightarrow  \tilde h^{-1} h\,,
\qquad {\rm for} \quad \tilde h \in {\rm H}\,.
\eqn\csec
$$
This H-gauge invariance is, of course, trivial since it derives from
the decomposition (\dec).  Interestingly, however, it can be made
non-trivial when $K \ne 0$, {\it i.e.,} when a non-trivial
quantization is considered.
To see this, let us first note that
the H-gauge invariance can still be observed
even after we
{}fix the $\SK$-gauge invariance.  Indeed,
since gauge fixing for the
$\SK$-invariance is achieved simply by decomposing $h$ using
a section $\tau : {\rm H}/\SK \mapsto {\rm H}$,
$$
h = \tau\,s\,,
\qquad {\rm with} \qquad \tau \in {\rm H}\,, \quad s \in
\SK\,,
\eqn\fsect
$$
and setting $s = 1$ subsequently,
the change of section
can still be performed by the slightly modified transformation,
$$
\s \longrightarrow \s \, \tilde h\,, \qquad
\tau \longrightarrow \tilde h^{-1}  \tau  \tilde s\,,
\qquad {\rm for} \quad \tilde h \in {\rm H}\,,
\eqn\msk
$$
where we choose $\tilde s = \tilde s(\tau, \tilde h) \in \SK$ (see
(\hsigma))
so that the transformation preserves the section $\tau$.
But since the transformation (\msk) amounts just to
the $\SK$-transformation $g \rightarrow g \tilde s$,
it leaves the path-integral (\cpi) invariant
thanks to the paramter quantization of $K$, hence the claimed
H-gauge invariance.

An important point to note is that
this transformation (\msk) does not involve a
coordinate transformation for $q$
(since $\s(q) \rightarrow \s'(q) = \s(q) \tilde h(q)$, see
Sect.\thinspace2.1), and
thus it can be interepreted
as an ordinary gauge transformation for the
background potential (\hcon) and the generalized spin
with $q$ fixed, under
which the total Lagrangian is invariant up to a
total time derivative.  (Note that this is a non-trivial gauge
transformation only for $K \ne 0$.)
In fact, the
H-gauge structure we have just observed is
a non-abelian, quantum mechanical version of
the abelian, functional gauge structure found by
Wu and Zee [\WZ]. This suggests that when our approach is applied
to Yang-Mills theory topological terms, and additional degrees of
{}freedom, can also arise there
{}from the  inequivalent quantizations
(for more detail, see [\MT]).
Another point worth mentioning is that the type of
Lagrangian (\llag) has been discussed by Balachandran
{\it et al.}\thinspace[\Balachandran] in an effort to incorporate
a Dirac monopole, non-relativistic spinning particle
and the Wong equations in gauge invariant formulations.
The discussion above shows that they all appear in the universal
{}framework in our approach based on
the reduction ${\rm G \rightarrow G/H}$.

We finish this section by noting the relation
between the singularity/topology of the H-connection and
the $\SK$-gauge invariance.
Recall that a section $\s = \s(q)$
can be taken only at the cost of introducing some
singularity in $q$
unless G is a direct product of G/H and H, {\it i.e.,}
unless the principal fibre bundle G is trivial.
This means that in a generic situation the H-connection is singular,
and it could be topologically non-trivial.
Clearly,
the locations of the singularity depend on the choice
of the section, and can be moved freely if a change of
section (\csec) is allowed.  As we have discussed above, this
is guaranteed by the $\SK$-symmetry at the quantum level, which is
ensured by the parameter quantization of $K$.
However, it must be stressed that the parameter quantization of
$K$ occurs {\it irrespective} of the non-triviality of the
principal bundle G --- it arises even when the bundle is trivial
and the H-connection is perfectly regular on G/H.
Such an example will be seen in the next section when we consider
the quantization on $S^3$.
\vfil\eject

\secno=6 \meqno=1

\noindent
{\bf 6. Physical implications of the inequivalant quantizations on $S^n$}
\medskip

\noindent
In this section we shall study the physical
implications of
inequivalent quantizations by quantizing on
spheres $S^n$, regarding them as the coset spaces
$S^n \simeq {\rm SO}(n+1)/{\rm SO}(n)$.
As we have seen in the previous sections, the hallmark of
inequivalent quantizations is the appearance of
the H-connection and
the generalized spin
given by a coadjoint orbit of H.
It will be therefore benefitial to start
with some basic remarks on the general properties of the
H-connection as well as of the coadjoint orbits of H.
After this we discuss inequivalent quantizations on the spheres
$S^n$ for $n =2$, 3 and 4 in detail\note{The general $S^n$ is
considered, though in less detail for the cases presented here,
in [\Ohnuki] which is essentially
based on the canonical group approach [\Isham] (see also
[\Tanimura]).}.
We shall find that the cases $S^2$, $S^4$ have
singular H-connections, which are identified
with a Dirac monopole and a BPST
(anti-)instanton, respectively.
In fact, a singular and topologically
non-trivial H-connection is expected in a generic case
where
G is not (as a topological space) a direct product of H and G/H.
The case
$S^3$ is non-generic in this sense since G $= {\rm SO}(4)$
is actually isomorphic to the direct product
${\rm SO}(4) \simeq {\rm SO}(3) \times S^3$ and hence the
H-connection is tolopogically trivial.
Nevertheless, the $S^3$ case is worth studying since it
serves as a simple example
where both the H-connection and the generalized spin emerge
simultaneously.
{}Furthermore, the generalized spin
turns out to be just the usual su(2) spin, {\it i.e.},
the conventional \lq non-relativistic'
spin degrees of freedom associated with the SO(3)
{}frame rotations of the \lq space'.
The case $S^4$ is even more interesting in that
the generalized spin consists of two su(2) spins, which
couple to a BPST
instanton and anti-instanton, chirally.
As in the $S^3$ case, the chiral su(2) spins admit the
interpretation that they are the two conventional \lq relativistic'
spins associated with the SO(4)
{}frame rotations of the \lq space-time'.

\bigskip
\noindent
{\it 6.1. Some generalities on the {\rm H}-connection}
\medskip

The canonical H-connection
has been studied earlier, {\it e.g.}, in the context of
the Kaluza-Klein theories and in
constructing topologically non-trivial Yang-Mills solutions [\BB].
{}For completeness
we wish to
recall here some of the basic properties of the H-connection
which we shall need later.
We begin by noting two salient features of the H-connection
which follow directly from the definition (\hconnection).
The first, which concisely
charcterizes the H-connection, is that
its curvature $F$ is constant
in the vielbein frame and
given by the structure constants
of the algebra $\g$.
To see this,
observe first
that the Maurer-Cartan equation for the left-invariant
1-form $\s^{-1} d\s$,
$$
d(\s^{-1}d\s) + (\s^{-1}d\s) \w (\s^{-1}d\s) = 0 \,,
\eqn\mc
$$
is equivalent to a pair of equations,
$$
dA + A\w A + e\w e|_\h = 0
\qquad \hbox{and} \qquad
de + e\w A + A\w e + e\w e|_\r = 0\,.
\eqn\pe
$$
One then finds that the first equation in (\pe), which is identical to
$$
{}F = - e \w e\vert_\h\,,
\eqn\cur
$$
implies that the curvature in the vielbein frame
$F = \frac{1}{2} F_{ab}^i T_i\, e^a \w e^b$ has
the components which are precisely the structure constants:
$$
{}F_{ab}^i = - f_{ab}^i\,.
\eqn\no
$$
Incidentally, we mention that the second equation in (\pe) can be
written as
$$
De=-e\w e|_\r\,,
\eqn\cc
$$
where $D$ is the covariant derivative.  This means that the vielbein
is convariantly constant $De = 0$ if G/H is a symmetric space.
(It is worth noting that for any coset G/H
one can define a Riemannian
spin connection 1-form satisfying the metricity and
the torsionless conditions, leading to a Riemannian manifold with
a constant curvature 2-form.  In particular, for a symmetric G/H the
spin connection is essentially equivalent to the H-connection.)

The second, more important, feature is that the H-connection
obeys the Yang-Mills equation on the coset space
G/H, that is,
$$
D^* F=0\,,
\eqn\ym
$$
where $D^* := - * D *$ is the adjoint operator of $D$.
We give a direct proof of this remarkable property
in Appendix D for a semisimple, compact G (an alternative proof can
be found, {\it e.g.}, in [\Laquer]).
In passing we wish to note that, for G/H =
SO($2m+1$)/SO($2m) \simeq S^{2m}$, it is known [\BB] that
if the H-connection is given in the spinor
representation of so($2m+1$) then it
can be decomposed into the two spinor representations of so($2m$) such
that one is self dual and the other anti-self dual (in the
generalized sense).  Thus,
in this case,
the fact that the H-connection being a solution of the Yang-Mills
equation on G/H follows directly from the Bianchi identity $DF = 0$.

Since the H-connection is necessarily singular
when the prinipal bundle G is non-trivial, the
construction of the H-connection provides --- in principle ---
a method for obtaining topologically non-trivial
solutions of the Yang-Mills equations on the coset G/H.
(In fact, the foregoing argument already
suggests that the H-connection would be a Dirac
monopole potential for ${\rm G/H = SO(3)/SO(2)} \simeq S^2$,
or a BPST instanton for
${\rm G/H = SO(5)/SO(4)} \simeq S^4$, which
we shall confirm soon.)  However, in practice,
we need to make some technical remarks on the situation where
such a topologically non-trivial solution actually occurs.
{}For simplicity let us
confine ourselves to the situation where G/H is an
even dimensional, symmetric
space.
In this situation one has
$F = - e \w e$ ({\it i.e.}, without
projection to $\h$)
and hence the original H-connection given by some
representation of $\g$ must have a vanishing Chern number as a whole,
$$
\tr (\overbrace{F \w \cdots \w F}^{m \rm\; times})
= (-1)^m \tr (\overbrace{ e \w \cdots \w e}^{2m \rm\; times}) = 0\,.
\eqn\ident
$$
This, of course, does not
mean that the H-connection must necessarily be topologically
trivial when it
is expressed in an irreducible representation of $\h$.

To be more explicit, let $\{\pi^\chi(T_i)\}$ be
the matrix form of a basis
$\{T_i\}$ in $\h$ in the irreducible representation $\chi$
of $\h$.  Then,
given a curvature $F$ in some representation of $\g$,
we may decompose it into irreducible representations
$\chi$ of $\h$ as
$$
{}F = \sum_\chi \oplus F^{\chi}, \qquad
\hbox{where}
\qquad
{}F^{\chi} := \pi^\chi(F)
= - \tfrac{1}{2} f_{ab}^i\, \pi^\chi(T_i)\, e^a \w e^b\,.
\eqn\no
$$
The Chern number for each $F^{\chi}$ is then proportional to
$$
\eqalign{
\tr (\overbrace{F^{\chi} \w \cdots \w F^{\chi}}^{m \rm\; times})
&= (-\tfrac{1}{2})^m \, \e^{a_1 b_1 \dots a_m b_m}\, f_{a_1 b_1}^{i_1}
   \cdots f_{a_m b_m}^{i_m}\, \cr
&\qquad\qquad \times \tr\bigl(\pi^\chi(T_{i_1}) \cdots
\pi^\chi(T_{i_m})\bigr)\,
\frac{\Omega}{\sqrt{{\rm det}\, \eta_{ab}}} \,,
}
\eqn\chno
$$
where
$\Omega = \sqrt{{\rm det}\,\eta_{ab}} \, e^{1} \w \cdots \w e^{2m}$ is the
volume form of the coset space G/H.
This decomposition may give us a non-trivial H-connection in some
representation $\chi$ of $\h$.  But if, in particular,
the coset space G/H is four
dimensional ({\it i.e.}, $m = 2$), and if $\h$ is simple and compact,
then for any $F^{\chi}$ the Chern number is zero.
Indeed, for such $\h$ every $\chi$ (apart from the
trivial one) is faithful and has the quadratic trace
proportional to the Killing form with the same sign.
{}From (\chno) this means that
the Chern numbers for all $F^{\chi}$ have the same sign,
but since
from (\ident) they must
add up to zero when summed over the $\chi$, they must be all
zero.  If, however, $\h$ is semisimple but
not simple, then $F^\chi$
may have a non-vanishing Chern number because such $\h$
admits non-faithful (but non-trivial) representations.
Actually, this is the case we shall encounter later when we consider the
quantization on $S^4$.

\bigskip
\noindent
{\it 6.2. Some generalities on the coadjoint orbits of \rm H}
\medskip

The coadjoint orbits of a group have also been well studied;
{}for instance, it is known that the orbits are K\"ahler manifolds,
among which the minimal dimensional ones are completely classified
{}for all classical groups, while
the maximal dimensional ones --- which arise for
regular semisimple $K$ and are just $O_K = {\rm H/T}$
with T being the maximal torus T of H --- are
{}flag manifolds (see, for example [\Woodhouse, \Perelomov, \AM]).
Although we do not need any of these
special properties of the coadjoint orbits later, we shall need
the following two.  The first is quite simple:
the variables $S_i = - \Tr (T_i h^{-1} K h)$
defined in (\gspin) describe the generalized spin.
This is based on the fact that they form
the \lq generalized spin algebra' in the reduced system, that is,
their Dirac bracket
(defined with respect to the second class
constraints $\varphi_k = (\phi_i, \xi_s)$ considered in
Sect.\thinspace5) is the algebra $\h$ itself:
$$
\{ S_i\,, S_j\}^* = f_{ij}^k \, S_k\,.
\eqn\salg
$$
To prove
this, note that $L_{O_K}$ would have been the only term
in the total Lagrangian (\llag)
if our reduction had been H/H (rather than G/H) where the
entire right-currents $\R_i$ are constrained to be $K_i$.
Obviously, in this reduction
the left-currents $\L_i$ in (\lr),
which reduce to $S_i$ on the constrained surface,
commute with the
constraints under Poisson bracket and thus are gauge invariant.
It then follows that the Dirac bracket
among $\L_i$ is equivalent to the Poisson bracket
in the reduced system ({\it i.e.},
on the second class constrained surface)
and hence we get (\salg)\note{In other words, just as we saw for the
spin example in Sect.\thinspace3,
the system of the coadjoint orbit $O_K$ of H can be
obtained simply by the reduction
${\rm H \rightarrow H/H}$ in our
generalized Dirac approach.  The same technique can be applied even
to loop groups LG, where one finds, {\it e.g.}, the model of
non-abelian chiral bosons as a coadjoint orbit LG/G [\TF].}.
Upon quantization, the Dirac bracket
(\salg) is replaced by the quantum commutator,
$$
[ \widehat S_i\,, \widehat S_j ] = i\, f_{ij}^k \, \widehat S_k\,.
\eqn\scom
$$
In view of the quantum
theory discussed in Sect.\thinspace4, the Hilbert space on which the
generalized spin is represented is given by
the highest weight
representation $\chi$ of H with $\chi(H_{\alpha_r}) = K_r$.

The second property we wish to note is
slightly more involved; it is
on the \lq space-time' nature of the generalized spin given by
the coadjoint orbit $O_K$.
Let us observe that the change of section (5.19)
induces, in addition to the gauge transformation (5.20) for the
H-connection, the following transformation for the vielbein,
$$
e = e^a T_a \longrightarrow \tilde h^{-1} e\, \tilde h =
e^a M_a^{\,\,b}(\tilde h) \, T_b \,,
\qquad{\rm with} \quad
M_a^{\,\,b}(\tilde h) := \eta^{bc} \Tr (\tilde h^{-1} T_a \tilde h \,T_c)\,,
\eqn\rot
$$
where $\{ T_a \}$ is a basis in $\r$, which specifies the
(vielbein) frame in
the tangent space defined at each point on the coset space G/H.
Since (\rot) leaves
the metric $g_{\alpha\beta}$ invariant, it is
an SO($n$) ($n = {\rm dim\,(G/H)}$) rotation of the vielbein
{}frame.  To put it differently,
on account of the reductive decomposition (A.6)
the complement $\r$ automatically furnishes a linear space
{}for a representation of the group H by the adjoint action (\rot),
producing the SO($n$) frame rotation.
In fact, we shall see that for $S^3$
the representation in the vielbein frame
is given by the adjoint of SU(2), that is, the
defining representation of SO(3), and hence provides
the totality of the frame rotation.
A similar situation occurs
{}for $S^4$ where again the representation in the vielbein frame
by the adjoint action (\rot)
provides the totality of the frame rotation SO(4).
Thus, combined with the response (\repspin)
under the change of section
in the sector $\chi$,
this observation enables us to determine the representation of
the generalized spin under the frame rotation in those cases.

\bigskip
\noindent
{\it 6.3. Quantizing on $S^n \simeq {\rm SO}(n+1)/{\rm
SO}(n)$}
\medskip

We are now in a position to
elaborate the physical consequences of inequivalent quantizations
by the examples of a particle moving on the
sphere $S^n$, realized as
the coset
SO($n+1$)/SO($n$), for $n = 2$, 3 and 4.
Our aim is to examine the role
played by the generalized spin, as well as to find
explicitly what the H-connections are.
We begin with the simplest case $S^2 \simeq$ SO(3)/SO(2).
\goodbreak
\medskip
\noindent
{\sl i) $S^2$ --- Dirac monopole:}
\smallskip

Let us choose for our basis
$
T_m = {{\s_m}\over{2i}}
$
for which
$
\tr (T_m\, T_n) = -\frac{1}{2}\, \d_{mn},
$
and take the orthogonal decomposition of $\g$ as
$$
\g = \h \oplus \r =
{\rm span} \{ T_3 \}
\oplus
{\rm span}\{T_1, T_2 \}\,.
\eqn\no
$$
Although the H-connection does emerge in this case,
no spin degrees of freedom appear
(since the coadjoint orbits of SO(2) are points).
Thus the total Lagrangian (5.15) reduces to
$$
L_{\rm tot} = \frac 1 2 g_{\alpha\beta}(q)\, \dot q^\alpha \dot q^\beta
- \Tr \bigl(K A_\alpha(q)\bigr)\, \dot q^\alpha \,,
\eqn\lagd
$$
where $K = j T_3$.
If we set the radius of $S^2$ to unity,
then from Appendix C we find $c_{\cs{sp}} = {1\over2}$, which means that
our metric in the vielbein frame is
$\eta_{ab} = \d_{ab}$.  The curvature $F$ of the H-connection is
just
$$
F = - {1\over2} [T_a, T_b]\, e^a \w e^b = - T_3\,\Omega,
\eqn\no
$$
with $\Omega = e^1\w e^2$ being the volume form of the unit sphere
$S^2$,
which implies that the H-connection is
a Dirac monopole potential on $S^2$ as expected.

It is reassuring to see that the Dirac condition
{}for the monopole charge follows directly
{}from the quantization condition of the parameter in $K$.
Indeed, since
the effective abelian H-connection
coupled to the particle is $ \bar A :=
- \Tr (K A)$ and hence
the effective curvature
$ \bar F = - j \,\Tr (T_3 F) = j \Omega$,
the monopole charge $g$
--- given by $(-{1\over2})$ times
the first Chern number $C_1$ --- is found to be
$$
g = - {1\over2} C_1 [\bar F]
= {1\over{4\pi}} \int_{S^2} \bar F
= {j\over{4\pi}}\int_{S^2} \Omega = j\,.
\eqn\no
$$
Then by the parameter quantization
$j \in \Z$ (see Sect.3)
we recover the Dirac condition for the monopole charge.
(Had we started with $S^2 \simeq$ SU(2)/U(1), then on account of
$2 j \in \Z$
we would have seen $g$ to be a multiple of $\frac12$ ---
the standard result for the monopole charge in that setting.)

\medskip
\noindent
{\sl ii) $S^3$ --- emergence of spin:}
\smallskip

The case $S^3$ is doubly exceptional; first, it is a group
manifold $S^3 \simeq {\rm SU(2)}$ by itself, and thus
one can quantize without
regarding $S^3$ as a coset, yielding a unique quantum theory
where neither the H-connection nor generalized spin occur.  Second,
even if one regards $S^3$ as the coset
SO(4)/SO(3), on account of
the direct product structure\note{
Another SO(n) which admits a similar direct product is
SO(8) $\simeq$ SO(7) $\times S^7$.}
of the space
${\rm SO(4)} \simeq {\rm SO(3)} \times S^3$ one finds
that the H-connection is topologically trivial (although its
curvature is non-vanishing).
Nonetheless, this example is very
interesting not only in that it provides
one of the simplest coset spaces where we find
both of the effects
--- the H-connection and the generalized spin --- but also in that
the generalized spin can be regarded as the
conventional su(2) spin attached to the particle.

To see this, we shall work in the defining representation
$\cs{def}$ of SO(4) with the basis
$\{T_{ij}\}$ given in (C.5).  Using
$T_i := - \frac12 \e_{ijk} T_{jk}$ and $T_a := T_{a 4}$,
we take the decomposition,
$$
\g = \h \oplus  \r
= {\rm span}\{ T_i \} \oplus {\rm span} \{ T_a \}\,,
\qquad i, \, a = 1, \,2, \,3\,.
\eqn\decs
$$
The basis $\{ T_i \}$ forms an
su(2) algebra $[T_i\,, T_j] = \e_{ijk}\, T_k$,
with respect which
the basis in $\r$ becomes a (spin 1) vector,
$$
[T_i\,, T_a] = \e_{iab}\,T_b \,.
\eqn\vect
$$
Note that from the argument given earlier the
variables $S_i$
{}form an su(2) algebra under the Dirac bracket,
$$
\{S_i\,, S_j\}^* = \e_{ijk}\, S_k \,.
\eqn\spin
$$
This is also consistent with the fact that the
Lagrangian $L_{O_K}$ for the coadjoint orbit becomes the usual
spin Lagrangian $L_{\rm spin} = - j \Tr T_3 (h^{-1} \dot h)$
discussed in Sect.\thinspace3.
{}Furthermore, since the basis $\{T_a\}$ forms a
vector (\vect), the adjoint action (\rot)
yields the usual adjoint representation of the group
SU(2), {\it i.e.}, the defining representation of
SO(3).  This implies that any SO(3)
rotation of the vielbein frame is realized by the adjoint action
(\rot), under which
the spin $S_i$ furnishes the representation
of total spin $j$ --- the conventional picture of
\lq non-relativistic' spin responding under the
SO(3) rotations in \lq space'.  Thus we see that
quantizing on
$S^3$ provides another example (besides Mackey's account of spin by
quantizing on $\Real^3$) in which
the particle acquires spin due to
the inequivalent quantizations.

On the other hand, the H-connection
is regular everwhere on $S^3$
since there exists a globally well defined section.
Thus this $S^3$ case serves as a concrete example where
the parameter quantization of $K$
(or the quantization of the coupling constants) is not linked to the
singular nature of the H-connection.  This is in contrast to
the generic situation where the two
--- the singularity of the H-connection
and the parameter quantization --- are closely linked ({\it
e.g.}, the quantization of a Dirac monopole charge
is  often derived by requiring the insensitivity of
the path-integral to the location of the
singularity).

{}Finally, we mention an ambiguity in the reductive
decomposition: in general the reductive decomposition is far from
unique, and the properties of the H-connection and the generalized
spin depend on the decomposition.  For example, in the present case
we may take, instead
of (\decs), the
decomposition of so(4) into two commuting su(2) subalgebras as
${\rm so(4)} = {\rm su(2) \oplus su(2)}$.
Then, using this, we will find that the
H-connection vanishes identically on $S^3$; only the spin degrees of
{}freedom appear but decoupled from the free particle moving on $S^3$.
Moreover, the relation in the group representation of H between the frame
rotation and the spin will be lost;
the frame rotation becomes completely independent from the
SU(2) group action of the spin.  Thus this new
reductive decomposition does not provide the conventional spin
in this respect.

\medskip
\noindent
{\sl iii) $S^4$ --- BPST instanton/anti-instanton and two chiral spins:}
\smallskip

The case $S^4$ allows for a more intriguing physical interpretation
of
the consequences of the inequivalent
quantizations; we shall see that the H-connection consists of
a BPST instanton and anti-instanton, and that the
generalized spin splits into two
su(2) spins coupled chirally to the
instanton
and anti-instanton, respectively.  To show this, we use
the spinor representation of so(5), and choose for our basis
$\{T_m\} = \{T_a, T_i\}$ with
$$
T_a = \cases{
{1\over{2i}}
\left(\matrix{
       & \s_a \cr
  \s_a &      \cr}
\right),
& for $a = 1, \, 2, \, 3$, \cr
{1\over2}
\left(\matrix{
  & - 1 \cr
1 &   \cr}
\right),
& for $a = 4$, \cr
}
\eqn\no
$$
and
$$
T_i = \cases{
{1\over{2i}}
\left(\matrix{
\s_i    &  \cr
     &  \s_i  \cr}
\right),
& for $i = 1, \, 2, \, 3$, \cr
{1\over{2i}}
\left(\matrix{
\s_{i-3}    &  \cr
     &  - \s_{i-3}  \cr}
\right),
& for $i = 4, \, 5, \, 6$, \cr
}
\eqn\hrep
$$
{}for which $\tr (T_m\,T_n) = - \d_{mn}$.
The decomposition is taken to be
$$
\g = \h \oplus \r =
{\rm span}\{ T_i \}
\oplus
{\rm span}\{ T_a \}.
\eqn\no
$$
Adopting the unit radius $S^4$, we find $c_{\cs{sp}} = 1$ from
Appendix C and hence the
metric in the vielbein frame is $\eta_{ab} = \d_{ab}$.

Recall that for any $m$
the spinor representation of so($2m+1$), when restricted to
so($2m$), is decomposed
into the two spinor representations of so($2m$).  In
the present case $2m = 4$
the two so(4) spinor representations given in (\hrep)
are reducible,
which of course is due to the direct sum structure
so(4) $=$ su(2) $\oplus$ su(2).
Thus it is already clear that we
have two su(2)-valued variables for our generalized spin.
In order to
treat the two su(2) separately, which we label
by su$(2)^+$ and su$(2)^-$, it is convenient to introduce
the chiral basis
$$
T^{+}_i = \frac12 (T_i + T_{i+3}) =
{1\over{2i}}
\left(\matrix{
\s_i    &  \cr
     &  0  \cr}
\right),
\qquad
T^{-}_i = \frac12 (T_i - T_{i+3}) =
{1\over{2i}}
\left(\matrix{
0    &  \cr
     &  \s_i  \cr}
\right),
\eqn\cbasis
$$
{}for $i = 1, \, 2, \, 3$, and thereby write
$h = h^+\, h^-$,
where $h^\pm$ belong to the exponential groups generated by the chiral
su$(2)^\pm$ in the basis (\cbasis).
Setting $K = j^+ T^+_3 + j^- T^-_3$, we find that
the Lagrangian for the coadjoint orbit of SO(4) consists of those for
the coadjoint orbits of the two su(2),
{\it i.e.}, for the two conventional spins,
$$
L_{O_K} = L_{\rm spin}^+ + L_{\rm spin}^- =
- j^+ \Tr T^+_3 \bigl((h^+)^{-1} \dot h^+\bigr)
- j^- \Tr T^-_3 \bigl((h^-)^{-1} \dot h^-\bigr)\,.
\eqn\no
$$
Moreover, from (\salg) we have the variables
$S_i^{\pm} = - j^\pm \Tr \bigl( T^\pm_i (h^{\pm})^{-1} T_3^\pm
h^{\pm}\bigr)$ forming
two commuting su(2) algebras under the Dirac bracket,
$$
\{ S^+_i\,, S^+_j \}^* = \e_{ijk}\, S^+_k\,, \qquad
\{ S^-_i\,, S^-_j \}^* = \e_{ijk}\, S^-_k\,, \qquad
\{ S^+_i\,, S^-_j \}^* = 0\,.
\eqn\no
$$

{}From the relations,
$$
[T^\pm_i\,, T_a ] = \frac12 \e_{iab} \, T_b  \pm \frac12 \d_{ia}\,
T_4\,, \qquad
[T^\pm_i\,, T_4 ] = \mp \frac12 \d_{ia}\, T_a\,,
\eqn\no
$$
it is also easy to see that
 the basis
$\{T_{\pm\pm} := T_1 \mp i T_2,\,
T_{\pm\mp} := T_3 \mp i T_4\}$ in the space $\r$ forms
a tensor product representation
$2^+ \otimes 2^-$
of spin $\frac12$ with respect to each chiral su$(2)^\pm$.
Accordingly, the adjoint action (\rot) of ${\rm H = SO(4)}$
amounts to the product ${\rm SU(2) \times SU(2)}$ transformation in
the vielbein frame.
However, taking into account the fact that
the spinor representations of SO(4)
are \lq double-valued' as a group
representation, we
conclude
that the \lq true group\rq\ described by this adjoint action
is $({\rm SU(2)} \times {\rm SU(2)})/\Z_2$, which
is the totality of the frame rotation SO(4).  (In other words, the
representation of ${\rm H = SO(4)}$ in the vielbein frame
is faithful and \lq single-valued', and thus the group action of H
has a one to one correspondence to
the SO(4) frame rotation.)
This is exactly the same situation
encountered in the Minkowski space version of the
{}frame rotation, where
the Lorentz frame rotation (that is, the action of the
proper, orthochronous Lorents group) is realized by
the group action of SL(2,$\C$) consisting of two
chiral SU(2) actions.
Thus we have seen that, on $S^4$, we recover the conventional,
two su(2) chiral
\lq relativistic' spins from the inequivalent quantizations.

Turning to the H-connection, we observe that
the su(2)-valued
H-connections $A^\pm$, defined by the decomposition
$A = A^+ + A^-$ in terms of the chiral basis (\cbasis),
couple to the chiral su(2) spins chirally,
$$
L_{\rm int} = \Tr (S^+ A^+_\alpha) \, \dot q^\alpha
              + \Tr (S^- A^-_\alpha) \, \dot q^\alpha\,.
\eqn\no
$$
We now show that these chiral H-connections $A^\pm$
are nothing but a BPST instanton and anti-instanton, respectively.
We do this by computing explicitly
the Chern number of each of the su(2)-valued
H-connection $A^\pm$; from (\chno) we get
$$
{}F^\pm \w F^\pm = \tfrac1 4 \e^{abcd}\, f_{ab}^i\, f_{cd}^j \,
T^\pm_i \, T^\pm_j \, \Omega\,,
\eqn\ff
$$
where $F^{\pm}$ are the curvatures corresponding to $A^\pm$ and
$\Omega = e^1 \w e^2 \w e^3 \w e^4$ is the volume form on $S^4$
with unit radius,
$\int_{S^4} \Omega = {{8 \pi^2}\over 3}$.
The structure constants appearing in (\ff) can be read off from the
commutation relations,
$$
[T_a\,, T_b] = \e_{abi}\, T_i
\qquad {\rm and} \qquad
[T_a\,, T_4] = \d_{a,i-3}\, T_i\,,
\eqn\corel
$$
{}for $a$, $b = 1$,  2, 3.
Then
the instanton number
{}for $F^\pm$ ---
given by $(-1)$ times the second Chern number $C_2$ ---
is evaluated to be
$$
n^\pm = - C_2 [ F^\pm ] =
- {1\over{8\pi^2}}\int_{S^4} \tr (F^\pm \w F^\pm)
= - {1\over{8\pi^2}}\cdot
  {{8\pi^2}\over 3}\cdot(\mp 3) = \pm 1\,,
\eqn\no
$$
which shows that $A^\pm$ is indeed a BPST instanton
(anti-instanton).  One can also confirm that $F^\pm$
satisfies the self dual (anti-self dual) equation:
$$
* F^+ = F^+, \qquad * F^- = - F^-.
\eqn\no
$$

We finish this section by remarking on
a techincal point underlying the
topological property of the H-connection.
We have argued in Sect.\thinspace6.1 that for a four dimensional
symmetric space G/H the H-connection must necessarily be
topologically trivial if $\h$ is a simple algebra.
This however does not apply to the above case;
so(4) is semisimple but not simple.  In fact,
the two su(2) representations given in the chiral basis (\cbasis)
are not faithful representations of so(4), which allows the
H-connection to escape from
the argument given there, although the total
sum of the two Chern numbers is zero as required in (\ident).

\vfil\eject

\secno=7 \meqno=1
\noindent
{\bf 7. Discussion}
\medskip

\noindent
In this paper we have seen that Dirac's formulation of
constrained quantization  has to be generalized in order to
recover the known quantizations on the coset space G/H by reducing
{}from the naive
quantization on G. In this generalization the classical first class
constraints $\R_i=0$ become, in the quantum theory, a mixed set of
constraints $\R_i=K_i$, where the constants $K_i$ label the
irreducible representations of H. We have developed a path-integral
account of this within which the quantization of the constants
$K_i$ follow from the residual $\SK$-invariance generated by the
{}first class subset of the constaints. This path-integral
{}formulation  clearly shows how the different reductions are
associated with
additional, compact degrees of freedom which we identify with
generalized spin, and how the effective dynamics is
described in terms of a minimal coupling to the H-connection.

The generalization we have presented here is not unique. Indeed it is
also possible to recover the quantizations on G/H, while still
keeping with the original constraint,  by modifying the first step in
the constrained quantization process [\LLA]; that is, to keep with
one reduction but to allow
{}for many quantizations on G by identifying it with the
coset space $({\rm G}\times{\rm H})/\widetilde{\rm H}$, where
$\widetilde{\rm H}$ is the diagonal subgroup of ${\rm G}\times{\rm
H}$. Then, appealing to Mackey's analysis of such a coset space, there
will now be many
inequivalent quantizations on G labelled by the irreducible
representations of $\widetilde{\rm H}\simeq{\rm H}$. The reduction
to the quantizations on G/H can now be accomplished by applying the
standard Dirac approach to these many quantizations on G.
Although
this procedure has ended up with the correct result, we do not
{}feel that it is a satisfactory constrained method since it
still involves the use of
vector-valued wave functions on G, hence obstructing
a continuous path-integral formulation.  Besides,
the reliance on a non-trivial quantization on
the extended configuration space cannot  be extended in any obvious
way  to field theory
(a fuller account of these points is presented in [\David]).

{}First class constraints generate symmetries. In these quantum
mechanical examples the symmetries are local in time, and are thus
the natural model for the gauge symmeries (local in space and time) found
in Yang-Mills theories.
The most startling aspect of our analysis is that we have seen that
these classical symmetries do not, and {\it should not},
always go through
to the quantum theory. Thus, contrary to our normal expectation,
anomalous behaviour is the norm when quantizing gauge theories! Of
course, in Yang-Mills theory coupled to chiral fermions, the naive
quantization is being used: it then being the matter fields that cause
the anomalous structure. Indeed, the chiral anomaly is probing the
topology of the Yang-Mills configuration space, and is, in fact,
reflecting the existence of non-contractable two-spheres in the
reduced configuration space [\Nelson]. However, due to the very
existence of these topological structures, we expect there to be
non-trivial quantizations possible such that the resulting anomalous
structure will be the sum of that comming from the chiral fermions
and that comming from the particular quantum sector one is working
in. That such topological structures lead to inequivalent
quantizations in Yang-Mills theory follows quite naturally from our
constrained formulation and will be presented elsewhere [\MT].
The resulting functional
connections, generalizing the H-connection encountered in this
paper, will be seen to generalize the results of [\WZ].

The situation here is, we feel, analogous to that found when
discussing the Aharonov-Bohm effect [\AB]. In this situation there is an
inaccessible region in the plane through which a flux is passing,
resulting in the charged particles picking up a phase when they pass
around that region. This phase is apparently well defined and corresponds
to the flux
through the system.
However, if the region through which the flux is passing is really
inaccessible  to the particles, then the configuration space is
topologically a circle and there will be inequivalent quantizations
labelled by a phase\note{The phase here is characterizing the
irreducible representations of the group of integers $\Z$; that is
we are identifying
$S^1=\Real/\Z$ and applying Mackey to this coset space. A similar
situation arises in gauge theories with the emergence of the
$\theta$-vacua
[\Jackiw].}.  Thus the observed phase is really the sum of
two terms, one representing the actual flux put into the system by
hand, the other determined by the particular quantum sector one is in.
The intriguing point in the chiral gauge theory example is then to
determine whether
it is possible to quantize such that the resulting
amomalies coming from the two sources cancel each other.

Extending this argument, we  note that there appears to be some connection
between the inequivalent quantizations described in this paper and
the geometric phase that has become popular in physics. Indeed, the
emergence of a monopole when quantizing on a
two-sphere is
reminiscent of the occurence of a monopole when an adiabatic
transport is performend in a  parameter space which is a
two-sphere [\Berry]. A non-abelian generalization of Berry's
phase is possible [\Wilczek] such that the parameter space is an
$n$-sphere, and the associated H-connection (in this case an
SO(n)-connection) emerges.  (For a  more general framework
based on the generalized coherent state path-integral, see
[\Kuratsuji].)
In these
discussions the naive quantization is always being used, so we would
again
expect that the resulting observed phases should also have a
decomposition into the naive term and a part labeling the quantum sector
of the theory.   Such a point of view will be developed elsewhere.

In conclusion, we have presented here a constrained description of
how inequivalent quantizations arise. What is missing is a
constrained account of why they arise (a possible account is given
in [\David]).  However, we hope
that the methods presented here will, at least, help to make the
important ideas of
Mackey seem more accessible, and indeed more relevant, to the general
physics community.
\bigskip
\noindent
{\bf Acknowledgements:}  We both wish to thank John Lewis and
Lochlainn O'Raifeartaigh for
their support in this work, and I.T.\thinspace wishes
to thank L\'aszl\'o Feh\'er for discussions.

\vfil\eject

\secno=0
\appno=1
\meqno=1

\noindent
{\bf Appendix A. Conventions}
\bigskip
\noindent
In this appendix we shall provide our conventions along with some
basic facts of Lie groups/algebras [\Barut, \Humphreys] used in the text.

\smallskip
\noindent
{\it Innerproduct.}
Let G be a semisimple group and $\g$ its Lie algebra.
Using some irreducible
representation $\chi$ of the algebra $\g$,
one can define in $\g$
an innerproduct $(\cdot\,,\cdot)$ by the
{}formula
$$
(A,B) = \Tr (AB)\,, \qquad {\rm where} \qquad
\Tr(AB) := - {1\over{c_\chi}} \tr (\pi^\chi(A) \pi^\chi(B))\,,
\eqn\ip
$$
where $\pi^\chi(A)$, $\pi^\chi(B)$ are the matrices
in the representation $\chi$ which correspond to $A$, $B \in \g$.
(When it is obvious, we often omit $\pi^\chi$ for brevity.)
{}For a simple group G
the (positive) constant $c_\chi$ is related to the
index of the representation, which is inserted to make
the innerproduct representation-independent.
Note however that the constant $c_\chi$ can be rescaled freely; only the
relative normalizations among representations are fixed.

\smallskip
\noindent
{\it Highest weight representation.}
In $\g$, or in the complex extension $\g_{\rm c}$ of $\g$,
one can choose
(with respect to some Cartan subalgebra) the
Chevalley basis $\{H_\alpha, E_{\pm\varphi} \}$ where $\alpha$
are simple roots and $\varphi$ are positive roots.
The basis has the relations
$$
[E_\alpha, E_{-\alpha}] = H_\alpha\,, \qquad [H_\alpha, E_\beta] =
K_{\beta\alpha} E_\beta\,,
\eqn\kw
$$
{}for simple roots $\alpha$, $\beta$, and $K_{\beta\alpha} =
\beta(H_\alpha) =
{{2\beta\cdot\alpha}\over{\vert\alpha\vert^2}}$ is the Cartan matrix.
To every dominant weight $\chi$
there exists an irreducible representation
--- highest weight
representation --- of $\g$ where the Cartan elements $H_\alpha$
are diagonal; in particular,
on the states $\ket {\chi, \mu}$
specified by the weights $\mu$ connected to the
the dominant weight $\chi$ (identified as the highest weight
in the representation) their eigenvalues are all integer:
$$
H_\alpha \ket {\chi, \mu} = \mu(H_\alpha) \, \ket{\chi,  \mu} \,, \qquad
{\rm with}\qquad
\mu(H_\alpha) = {{2\mu\cdot\alpha}\over{\vert\alpha\vert^2}}
              \in {\Bbb Z}\,.
\eqn\hwr
$$
On account of this, we can use the dominant weight $\chi$ (or the set of
integers $\chi(H_{\alpha_r})$ for $r = 1, \ldots, {\rm rank\,G}$)
to label the irreducible
representation\note{In the text we  consider highest weight
representations $\chi$ of a subalgebra $\h$ of $\g$, or of
its group H for which we use the same notation.}.
The integral property (\hwr) of $H_\alpha$ leads to the
periodicity in the exponential mapping defined in
the universal covering group
$\widetilde {\rm G}$ of G ({\it i.e.}, the
simply connected group which shares the same algebra $\g$),
$$
e^{2 n \pi i H_\alpha} = 1\,, \qquad \hbox{for} \quad
n \in \Z\,.
\eqn\perd
$$
{}For a non-simply connected group G the periodicity is
different from (\perd); it is mulitplied by a factor
determined by the discrete normal subgroup
N of $\widetilde {\rm G}$ for which
G $\simeq \widetilde{\rm G}/{\rm N}$.
{}For instance, for ${\rm Spin}(n) = \widetilde{{\rm SO}}(n)$
we have $n \in \Z$ but for ${\rm SO}(n) \simeq
{\rm Spin}(n)/\Z_2$ we find $2n \in \Z$
(${\rm Spin}(3) = {\rm SU}(2)$).

\smallskip
\noindent
{\it Reductive decomposition}.
Given a closed subgroup H of G
with its Lie algebra $\h$,
we take an orthogonal decomposition of $\g$,
$$
\g=\h\oplus\r\,,
\eqn\od
$$
where $\r=\h^\perp$ is the orthogonal complement of $\h$ to $\g$
with respect to the innerproduct; $(\h, \r) = 0$.
This is, in fact, a reductive decomposition, {\it i.e.,}
$$
[\h,\r]\subset\r\,,
\eqn\red
$$
since $[\h,\h]\subset\h$ and the
orthogonality imply
$0 = ([\h,\h], \r) = (\h, [\h,\r])$ and hence the relation
(\red).
If, in addition, we have
$$
[\r,\r]\subset\h\,,
\eqn\sym
$$
then the coset space G/H, given by the left cosets $\{ g{\rm H}\,
\vert\, g\in {\rm G}\}$, is a symmetric space.

We shall denote bases of the spaces by
$$
\eqalign{
\g &= \hbox{span} \{T_m\}\,, \cr
\h &= \hbox{span} \{T_i\}\,, \cr
\r &= \hbox{span} \{T_a\}\,, \cr
}
\qquad
\eqalign{
m &= 1, \ldots, \dim{\rm G}\,, \cr
i &= 1, \ldots, \dim{\rm H}\,, \cr
a &= 1, \ldots, \dim{\rm (G/H)}\,. \cr
}
\eqn\conv
$$
As usual, any $A
\in \g$ can be expanded as $A = A^m T_m$ and one can raise or lower
the indices by using the (flat) metric
$\eta_{mn} := (T_m, T_n)$ and its inverse $\eta^{mn}$ with
$\eta^{ml} \eta_{ln} = \d^m_n$.
We use the structure constants $f^l_{mn}$ defined by
$[T_m, T_n] = f_{mn}^l \, T_l$; in more detail,
$$
[T_i,T_j] = f_{ij}^k\,T_k\,, \qquad
[T_i,T_a] = f_{ia}^b\,T_b\,, \qquad
[T_a,T_b] = f_{ab}^c\,T_c + f^i_{ab}\,T_i\,.
\eqn\sc
$$
If G/H is a symmetric space we have $f^c_{ab}=0$.
If, on the other hand, G is compact, then
the Killing form, defined by
the matrix trace of the adjoint representation, is negative
definite and hence
we can find a basis where the metric
takes the form
$\eta_{mn} = \d_{mn}$.  Using this basis, we can ignore
the differences between
upper and lower indices, and the structure constants
are totally antisymmetric in the three indices.  In this paper we
consider
only semisimple, compact groups for H, and
the coset G/H = SO($n+1$)/SO($n$),
which is our main concern in Sect.\thinspace6, is a symmetric space.

\vfill\eject


\secno=0
\appno=2
\meqno=1

\noindent
{\bf Appendix B. Free particle on the group manifold {\rm G}}
\bigskip
\noindent
In this appendix we shall provide a Hamiltonian description
of a free particle moving on the group manifold G.

In general,
a Hamiltonian system is defined by the triple
$(M, \{\cdot\,,\cdot\}, H)$
where $M$ is the phase space,
$\{\cdot\,,\cdot\}$ the Poisson bracket and $H$ the Hamiltonian.
{}For our purpose we take for $M$ the cotangent bundle of the group G,
$$
M = T^* {\rm G} = {\rm G} \times \g^* \simeq {\rm G} \times \g =
\{ (g, \R)\, \vert\,\, g \in {\rm G}, \, \R \in \g \} \,.
\eqn\pha
$$
The Poisson bracket can be given from the symplectic 2-form
$$
\omega = - d\, \Tr \R (g^{-1} dg)
= - d\, \bigl\{ \R_m (g^{-1} dg)^m \bigr\} \,,
\eqn\symp
$$
and our Hamiltonian is
$$
H = \tfrac 1 2 \Tr \R^2 = \tfrac 1 2 \eta^{mn} \R_m \R_n\,.
\eqn\ham
$$

In order to derive
the Poisson bracket
{}from the symplectic 2-form,
we first recall that to every function $f = f(g,\R)$ on $M$
there exists a corresponding Hamiltonian vector field $X_f$
satisfying
$$
i_{X_f} \omega := \omega ( X_f,\, \cdot\, ) = - d f\,,
\eqn\inc
$$
where $i_{X_f}$ is the contraction with the vector field $X_f$.
By using the left-invariant
1-forms $(g^{-1} dg)^m$
and their duals $Y_m$, the left-invariant vector fields
for which $\langle (g^{-1} dg)^m\,, Y_n \rangle = \d^m_n$,
one has
$$
d f = Y_m f \cdot (g^{-1} dg)^m
+ {{\pa f}\over{\pa \R_m}}\, d\R_m\,,
\eqn\df
$$
and hence from
(\symp) and (\inc) one finds
$$
X_f = - {{\pa f}\over{\pa \R_m}}\, Y_m +
\bigl( Y_m f + f_{mn}^l {{\pa f}\over{\pa \R_n}}\, \R_l \bigr) \,
{{\pa}\over{\pa \R_m}}\,.
\eqn\vf
$$
Then the Poisson bracket between the two functions $f$, $h$ on $M$
is given by
$$
\{f\,,\, h\} := \omega(X_h, X_f)
= {{\pa f}\over{\pa \R_m}}\, Y_m h - {{\pa h}\over{\pa \R_m}}\, Y_m f
+ f_{mn}^l {{\pa f}\over{\pa \R_m}}\,{{\pa h}\over{\pa \R_n}}\, \R_l\,.
\eqn\pb
$$
In particular, we have the following fundamental Poisson bracket,
$$
\{\R_m \,,\, \R_n\} = f_{mn}^l \, \R_l\,,
\qquad
\{\R_m \,,\, g\} = Y_m g\,.
\eqn\fpb
$$
Under the use of some representation $\chi$, the left-invariant vector
{}fields admit the explicit form
$Y_m = (T_m)_{ij} g_{ki} {{\pa}\over{\pa g_{kj}}}$.  Using this, one
{}finds that the second relation in (\fpb) becomes
$ \{\R_m \,,\, g_{ij}\} = (g T_m)_{ij} $, which confirms that the
\lq right-currents' $\R_m$, or equivalently the
left-invariant vector fields $Y_m$, are the generators of the right
translation.
It is also easy to
see that the \lq left-currents' $\L_m$ defined by
$\L = -g \R g^{-1}$ satisfy
$$
\{\L_m \,,\, \L_n\} = f_{mn}^l \, \L_l\,,
\qquad
\{\L_m \,,\, g\} =- Z_m g\,,
\eqn\fpbr
$$
where $Z_m$ are the right-invariant vector fields dual to the
right-invariant 1-forms $(dg\, g^{-1})^m$, {\it i.e.,}
$\langle (dg\, g^{-1})^m\,, Z_n \rangle = \d^m_n$.
Again, by using some
representation $\chi$ we have
$Z_m = (T_m)_{ij} g_{jk} {{\pa}\over{\pa g_{ik}}}$ and hence
the second relation in (\fpbr) reads
$ \{\L_m \,,\, g_{ij}\} =- (T_m g)_{ij} $, implying that
the left-currents generate the left translation.
Of course, the two generators commute:
$$
\{\R_m \,,\, \L_n\} = 0\,.
\eqn\com
$$

{}From (\ham) and (\fpb) the Hamilton equations read
$$
\dot g = \{ g\,,\, H \} = - g \R, \qquad
\dot \R = \{ \R\,,\, H \} = 0\,,
\eqn\hem
$$
where dots denote differentiation with respect to time $t$.
Combining the two equations in (\hem) we get
$$
{{d}\over{dt}}(g^{-1} \dot g) = 0\,.
\eqn\em
$$
To see that (\em) in fact
describes geodesic
motion of a particle moving on the group manifold G,
let us first recall that
a natural Riemannian metric on the group G is given by
the formula
$$
ds^2 = g_{\mu\nu}\, dx^\mu \otimes dx^\nu
:= \Tr (v \otimes v)\,,
\eqn\metg
$$
where $\{ x^\mu \}$, $\mu = 1, \ldots, N$ ($N = \dim$G) is a
coordinate on G, and
$v := g^{-1}dg$ is the vielbein defined from the left-invariant
1-form.
Expanding the
vielbein $v = v^m\, T_m = v_\mu\, dx^\mu = v^m_{\,\,\mu} \, dx^\mu
\, T_m$, one finds that
the metric in (\metg) reads
$g_{\mu\nu} = \eta_{mn}\, v_{\,\,\mu}^m\, v_{\,\,\nu}^n$,
where $\eta_{mn}$ is the flat metric ({\it i.e.}, the metric in the
vielbein frame) defined in Appendix A.  Using the metric, it is now
easy to rewrite (\em) in the desired form
$$
\ddot x^\mu + \Gamma^\mu_{\nu\lambda}\, \dot x^\nu \dot x^\lambda =
0\,,
\eqn\geo
$$
with the Levi-Civita connection,
$$
\Gamma^\mu_{\nu\lambda} = \frac 1 2 g^{\mu\rho}( \pa_{\nu}
g_{\rho\lambda}
+ \pa_\lambda g_{\rho\nu} - \pa_\rho g_{\nu\lambda})\,.
\eqn\levciv
$$

\vfill\eject


\secno=0
\appno=3
\meqno=1

\noindent
{\bf Appendix C. Metric on the coset {\rm G/H}}
\bigskip
\noindent
In this appendix we shall give a brief account of the metric
on the coset space G/H [\Gilmore].
Our purpose here is to provide
a simple procedure to determine the
constant $c_\chi$, which allows us
to evaluate explicitly the topological (Chern) numbers
of the H-connections in the text.

To furnish
a Riemannian metric on G/H
where G is a semisimple, compact group,
we first note that, given a (local) section $\s:$
G/H $\mapsto$ G in G,
we can decompose $g \in$ G as
$$
g = \s \, h\,, \qquad {\rm where} \qquad \s \in {\rm G}\,,
\quad h \in {\rm H}\,.
\eqn\sect
$$
Using the section, we consider
the 1-form $\s^{-1}d\s$ and its decomposition
with respect to $\g = \h \oplus \r$,
$$
\s^{-1} d\s = A + e\,, \qquad
{\rm where} \qquad
A := \s^{-1}d\s \vert_\h\,, \quad
e := \s^{-1}d\s \vert_\r\,.
\eqn\dec
$$
The 1-form $A$ is the H-connection and is discussed in detail in the
text, while the 1-form
$e$ is regarded as a vielbein.
As we did on the group manifold G, we define
a metric on the coset manifold G/H by
$$
ds^2 = g_{\alpha\beta} \, dq^\alpha \otimes dq^\beta
:= \Tr (e \otimes e)\,,
\eqn\metgh
$$
where $\{q^\alpha \}$, $\alpha = 1, \ldots, n$
($n = \dim$(G/H)), is a set of local coordinates on
G/H.  Expanding the vielbein
$e = e^a \, T_a = e_\alpha\,
dq^\alpha = e^a_{\,\,\alpha}\, dq^\alpha\, T_a$,
one finds the metric
$
g_{\alpha\beta} = \eta_{ab}\, e_{\,\,\alpha}^a\, e_{\,\,\beta}^b,
$
where $\eta_{ab} := (T_a, T_b)$ is the (flat)
metric defined by
the restriction of the metric $\eta_{mn}$ to the subspace $\r$.
Note that
the left translation $g \rightarrow \tilde g g$ with $\tilde g \in
{\rm G}$
induces a transformation
$\s(q) \rightarrow \s(\tilde g q) = \tilde g \s(q)
h^{-1}_\s(\tilde g, q)$  where $h_\s\in{\rm H}$
(see (\hsigma)), and hence
the vielbein $e$
undergoes the SO($n$) transformation $e \rightarrow h_\s e
h^{-1}_\s$ in the vielbein frame, leaving the metric
$g_{\alpha\beta}$ invariant.
Once we have a vielbein (or metric) on G/H the volume form is defined
as
$$
\Omega = \sqrt{{\rm det}\, \eta_{ab}}\, e^1 \w e^2 \w \cdots \w e^n
  = \sqrt{{\rm det}\, g_{\alpha\beta}} \, dq^1 \w dq^2 \w \cdots \w
dq^n\,.
\eqn\vol
$$

Let us specialize to the case G/H = SO($n+1$)/SO($n$) $\simeq
S^n$.  Working in the defining representation
$\cs{def}$ of so($n+1$), we
shall choose a basis $\{ T_{jk} \}$, $j$, $k = 1, \ldots, n+1$,
by
$$
(T_{jk})_{lm} := \d_{jl}\, \d_{km} - \d_{jm}\, \d_{kl},
\eqn\basis
$$
which has the trace,
$$
\tr ( T_{jk} T_{j'k'} ) =  - 2 \, \d_{j j'} \, \d_{k k'} \,.
\eqn\normal
$$
Defining $T_a = T_{a, n+1}$ for $a = 1, \ldots, n$,
we have the orthogonal decomposition
$$
\g = \h \oplus \r
= {\rm span}\{T_{ab} \} \oplus {\rm span} \{ T_a \}\,,
\qquad a, \, b = 1, \ldots, n.
\eqn\ordec
$$
If we choose for the decomposition (\sect) the section
$\s = e^{\theta^a T_a}$, we find that
$$
\s =
\left(\matrix{
\cos \sqrt{\theta\theta^t}   &  \dfrac{\sin
\vert\theta\vert}{\vert\theta\vert} \theta \cr
- \dfrac{\sin \vert\theta\vert}{\vert\theta\vert} \theta^t
     &  \cos \vert\theta\vert \cr}
\right),
\eqn\mat
$$
where we have used the column vector $\theta$ with
$\theta^t = (\theta^1, \ldots, \theta^n)$ and $\vert\theta\vert =
\sqrt{\theta^t \theta} = \sqrt {\sum_{a} (\theta^a)^2}$.
Then we find that the new coordinate
$q = (q^1,\ldots, q^{n+1})$ with
$$
q^a := \theta^a {{\sin \vert\theta\vert}\over{\vert\theta\vert}},
\quad a=1, \ldots, n, \qquad
q^{n+1} := \cos \vert\theta\vert,
\eqn\cd
$$
satisfies the condition $\sum_{a=1}^{n+1} (q^a)^2 = 1$,
and thus it serves as
a coordinate on $S^n$ with unit radius.

The metric on the coset is given by (\metgh), but in order to
specify the
radius $r$ of $S^n$ to be unity for the coset space we must determine
the constant $c_\chi$ used in the innerproduct
appropriately.  A practical way for this
is the following.
Take the two points, the north pole $q_{_N}$ and the south pole
$q_{_S}$
on $S^n$.
In terms of the coordinate $q$
these points may be taken to be
$$
q_{_N} = (0,0,\ldots,0, 1),\qquad
q_{_S} = (0,0,\ldots, 0, -1)\,.
\eqn\ns
$$
{}From (\cd) a great circle
connecting the two points is given by $\theta^1(t) = \pi t$,
$\theta^a(t) = 0$ for $a \neq 1$ $(0 \leq t \leq 1)$; indeed we have
$q(t) := q(\theta(t)) = (\sin \pi t, 0, \ldots, 0,
\cos \pi t)$ and hence $q(0) = q_{_N}$, $q(1) = q_{_S}$.
Accordingly, we get $\s(q(t)) = e^{\pi t T_1}$ and thus the
vielbein is just $e = \pi dt \,T_1$.  Since the distance between the
two points is evaluated from (\metgh) as
$$
d(q_{_N}, q_{_S}) := \int_{q_{_N}}^{q_{_S}} ds = \pi\,
\sqrt{- c_{\cs{def}}^{-1} \tr T_1^2 } \int_0^1 dt \,,
\eqn\dist
$$
and since this must be $\pi$
{}for $S^n$ with unit radius,
we deduce that $c_{\cs{def}} = 2$ for the defining
representation on account of the
normalization (\normal).

Let us specialize further to the $n = 2m$ situation;
G/H = SO($2m+1$)/SO($2m$) $\simeq S^{2m}$.
In this case, from
the indices of the representations, {\it i.e.,}
{}from the fact that the quadratic trace
in the adjoint (spinor) representation of
so($2m+1$) is $(2m-1)$
times ($2^{m-3}$ times) that of the defining representation,
we find
$$
c_{\cs{def}} = 2\,, \qquad
c_{\cs{ad}} = 2(2m-1)\,, \qquad
c_{\cs{sp}} = 2^{m-2}\,.
\eqn\cvalue
$$
If we wish to have a generic radius $r$, we just rescale those $c_\chi$ by
$\frac1{r^2}$; {\it e.g.}, we have $c_{\cs{def}} = \frac2{r^2}$.

\vfill\eject


\secno=0
\appno=4
\meqno=1

\noindent
{\bf Appendix D. {\rm H}-connection as a Yang-Mills
connection on} G/H
\bigskip
\noindent
We wish to show that,
{}for a semisimple, compact G,
the H-connection $A$
is in fact a solution of the Yang-Mills equation on the coset space
G/H, that is, the curvature $F := dA + A \w A$ satisfies
$$
D^* F=0\,,
\eqn\ym
$$
where $D^* := - * D *$ is the adjoint operator of $D$.
The proof is straightforward:

\smallskip
\noindent
{\it Proof.} From the definition of the Hodge-star $*$,
$$
*(e^{a_1}\w\cdots\w e^{a_p})
  =\frac1{(n-p)!}\,
\e^{a_1 \ldots a_p}_{\phantom{a_1\ldots a_p} a_{p+1} \dots a_n}
 e^{a_{p+1}} \w\cdots\w e^{a_n}\,,
\eqn\hodge
$$
we have for $D*F = d*F + [A, *F]$,
$$
\eqalign{
(D*F)^i=-\frac1{(n-3)!2}&f^i_{ab}\, \e^{ab}_{\phantom{ab}a_3\dots a_n}
d(e^{a_3})\w e^{a_4}\w\cdots\w e^{a_n}\cr
&-\frac1{(n-2)!2}f^i_{jk}f^k_{ab}\, \e^{ab}_{\phantom{ab}a_3\dots a_n}
A^j\w  e^{a_3}\w\cdots\w e^{a_n}\,.
}
\eqn\basic
$$
Using
the Maurer-Cartan equation
$
De=-e\w e|_\r
$
in (\cc) the first term in (\basic) becomes
$$
d*F^i = \frac1{(n-3)!4} f^i_{ab}
(f^{a_3}_{cd} + 2 f^{a_3}_{jd}\,A^j_c)\,
\e^{ab}_{\phantom{ab}a_3\dots a_n}
e^c\w e^d\w e^{a_4}\w\cdots\w e^{a_n}\,,
\eqn\fr
$$
where we have expanded $A^j=A^j_c\,e^c$ in the vielbein frame.
Then applying $*$ we get
$$
\eqalign{
*d*F^i
& = \frac1{(n-3)!4} f^i_{ab}
(f^{a_3}_{cd} + 2 f^{a_3}_{jd}\,A^j_c)\,
\e^{ab}_{\phantom{ab}a_3\dots a_n}
\e^{cda_4\dots a_n}_{\phantom{cda_4\dots a_n}e}e^e \cr
&=\tfrac12 (-1)^{n-1} f^i_{ab}f^e_{ab}\,e^e
    + (-1)^{n-1}(f^b_{ej}f^a_{bi}+f^b_{ie}f^a_{bj})A^j_ae^e\,,
}
\eqn\first
$$
where we have used a basis in which the metric in the vielbein
{}frame takes the form
$\eta_{mn} = \d_{mn}$, which is possible because of the compactness
of the group G (see Appendix A).
On the other hand, applying $*$ to the second term in (\basic) we
obtain
$$
*[A,*F]^i
= - \frac1 {(n-2)!2}
{}f^i_{jk}f^{k}_{ab}\,A^j_c \,
\e^{ab}_{\phantom{ab}a_3\dots a_n}
\e^{ca_3\dots a_n}_{\phantom{ca_3\dots a_n}e}e^e
= (-1)^{n-1} f^i_{jk}f^k_{ae}\,A^j_ae^e\,.
\eqn\second
$$
Combining (\first) and (\second) we find
$$
(-1)^{n}(D^*F)^i
= \tfrac12f^i_{ab}f^e_{ab}e^e+
 (f^b_{ej}f^a_{bi}+f^b_{ie}f^a_{bj}+f^i_{jk}f^k_{ae})A^j_ae^e\,.
\eqn\third
$$
By using the Jacobi identity and
the orthogonality of the decomposition (A.4),
$$
{}f^j_{ab}f^k_{ji}+f^c_{ia}f^k_{cb}+f^c_{bi}f^k_{ca}=0\,,
\qquad {\rm and} \qquad
{}f^b_{ea} f^a_{ib} = \tr (\pi^{\cs{ad}}(T_e) \pi^{\cs{ad}}(T_i)) = 0\,,
\eqn\jacobi
$$
we obtain the Yang-Mills equation (\ym). {\it Q.E.D.}

\vfill\eject

  \vfill\eject\immediate\closeout\reffile
  \centerline{{\bf References}}\bigskip\frenchspacing%
  \input refs.tmp\vfill\eject\nonfrenchspacing


\bye